\documentclass[aoas]{imsart}

\RequirePackage{amsthm,amsmath,amsfonts,amssymb}
\RequirePackage[authoryear]{natbib}
\RequirePackage[colorlinks,citecolor=blue,urlcolor=blue]{hyperref}
\RequirePackage{graphicx}

\startlocaldefs
\usepackage[anythingbreaks]{breakurl}
\usepackage{color}

\newcommand{\cB}{\ensuremath{\mathcal{B}}}
\newcommand{\cD}{\ensuremath{\mathcal{D}}}
\newcommand{\mc}[1]{\ensuremath{\mathcal{#1}}}

\newcommand{\EE}{\mathbb{E}}
\newcommand{\Bern}{\ensuremath{\mathrm{Bernoulli}}}
\newcommand{\SPR}{\ensuremath{\mathrm{SPR}}}

\renewcommand{\Pr}{\mathbb{P}}

\endlocaldefs

\begin{document}

\begin{frontmatter}
\title{ALPHA: Audit that Learns from Previously Hand-Audited Ballots}
\runtitle{ALPHA Risk-Limiting Audits}


\begin{aug}
\author[A]{\fnms{Philip B.}~\snm{Stark}\ead[label=e1]{stark@stat.berkeley.edu} 
}
\address[A]{Department of Statistics, University of California, Berkeley\printead[presep={,\ }]{e1}}
\end{aug}

\begin{abstract}
  A risk-limiting election audit (RLA) offers a statistical guarantee: if the reported electoral outcome
     is incorrect, the audit has at most a known maximum chance (the risk limit)
     of not correcting it before it becomes final.
     BRAVO \citep{lindemanEtal12}, based on Wald's sequential probability ratio test for the Bernoulli parameter, 
     is the simplest and most widely tried method for RLAs, but it has limitations.
     It cannot accommodate sampling without replacement or stratified sampling, which can improve efficiency 
     and are sometimes required by law.
     It applies only to ballot-polling audits, which are less efficient than comparison audits.
     It applies to plurality, majority, super-majority, proportional representation, and instant-runoff voting (IRV, using RAIRE \citep{blomEtal18}), 
     but not to other social choice functions for which there are RLA methods.
     And while BRAVO has the smallest expected sample size among sequentially valid ballot-polling-with-replacement
     methods when the reported vote shares are exactly correct, it can require arbitrarily large samples
     when the reported reported winner(s) really won but the reported vote shares are incorrect. 
     ALPHA is a simple generalization of BRAVO that (i)~works for sampling with and without replacement, with
     and without weights, with and without stratification, and for Bernoulli sampling; (ii)~works not only for ballot polling but also for ballot-level comparison, batch polling, and batch-level comparison audits; 
     (iii)~works for all social choice functions covered by SHANGRLA \citep{stark20}, 
     including approval voting, STAR-Voting, proportional representation schemes such as D'Hondt and Hamilton, IRV, Borda count, and all scoring rules;
     and (iv)~in situations where both ALPHA and BRAVO apply, requires smaller samples than BRAVO
     when the reported vote shares are wrong but the outcome is correct---five orders of 
     magnitude in some examples.
     ALPHA includes the family of betting martingale tests in RiLACS \citep{waudby-smithEtal21}, with a different betting
     strategy parametrized as an estimator of the population mean and explicit flexibility to accommodate sampling
     weights and population bounds that change with each draw.
     A Python implementation is provided.
\end{abstract}

\begin{keyword}
\kwd{risk-limiting audit}
\kwd{elections}
\kwd{supermartingale test}
\kwd{Ville's Inequality}
\kwd{SHANGRLA}
\end{keyword}

\end{frontmatter}


\maketitle

\section{Introduction}
A risk-limiting audit (RLA) is a procedure that has a known minimum probability of correcting the reported outcome of an election
contest if the reported outcome is wrong.
The risk limit of an RLA is the maximum chance that the RLA will not correct the electoral outcome, if the outcome is wrong.
The \emph{outcome} means the political outcome---who or what won---not the numerical vote tallies, which are practically impossible
to get exactly right.
An RLA requires a trustworthy record of the validly cast votes:\footnote{%
Generally, the record is a set of validly cast hand-marked paper ballot cards that has been kept demonstrably secure.
Machine-marked ballot cards cannot be considered a trustworthy record of voter intent.
See \citet{appelEtal20,appelStark20,starkWagner12}.
}
a manual count of those records is the recourse to correct
wrong outcomes.
Establishing whether the record of votes is trustworthy prior to conducting a risk-limiting audit is 
generically called a \emph{compliance audit} 
\citep{starkWagner12,appelStark20}.
RLAs are recommended by the National Academies of Science, Engineering, and Medicine \citep{nas18},
the American Statistical Association \citep{asa10}, and other groups concerned with election integrity.
As of this writing, RLAs are authorized or required by law in fifteen U.S.\ states and have been piloted in 
roughly a dozen U.S.\ states and in Denmark.

BRAVO \citep{lindemanEtal12} is a particularly
simple method to conduct an RLA of plurality and supermajority contests.
It relies on sampling ballot cards\footnote{%
In general, a ballot is comprised of one or more \emph{ballot cards}, each of which contains some of the contests
a given voter is eligible to vote in.
Many countries and some U.S.\ states have one-card ballots, but many U.S.\ states routinely
have ballots that comprise two or more
ballot cards.
}  
uniformly at random with replacement from all ballot cards validly cast in the contest.
\citet{starkTeague14} showed how BRAVO can be used to audit proportional representation schemes such as D'Hondt.
\citet{blomEtal18} showed how BRAVO can be used to audit instant-runoff voting (IRV), a form of ranked-choice voting.
BRAVO is based on Wald's \citep{wald45} sequential probability ratio test (SPRT) of the simple hypothesis 
$\theta = \mu$ against a simple alternative $\theta = \eta$ from IID $\Bern(\theta)$ observations.
(A $\Bern(\theta)$ random variable takes the value $0$ with probability $1-\theta$ and the value $1$ with probability $\theta$;
its expected value is $\theta$.)
Because it requires IID $\Bern(\theta)$ observations, BRAVO is limited to \emph{ballot-polling} audits and to using samples drawn with 
replacement, both of which limit efficiency and applicability.
(A ballot-polling audit involves manually interpreting randomly selected ballots, but does not use the voting system's
interpretation of individual ballot cards or groups of ballot cards---just the reported outcome.
As discussed below, \emph{comparison} audits, which compare the voting system's interpretation of ballot cards to
manual interpretations of the same cards, can be more efficient.)

To audit a plurality contest with BRAVO involves using the SPRT to test a number of hypotheses:
for each reported winner $w$ and each reported loser $\ell$, let $\theta_{w\ell}$ be the 
the conditional probability that a ballot selected at random with replacement
from all ballot cards validly cast in the contest shows a valid vote for $w$, given that it shows a valid vote either for $w$ or for $\ell$,
and let $\eta_{w\ell}$ be the number of votes reported for $w$, divided by the total votes reported for $w$ and $\ell$ combined.
For every  $(w, \ell)$ pair,
BRAVO tests the hypothesis $\theta_{w\ell} = 1/2$ against the alternative $\theta_{w\ell} = \eta_{w\ell}$.
No multiplicity adjustment is needed because the audit proceeds to a full hand count unless \emph{every} null hypothesis
is rejected.

BRAVO for a supermajority contest can be simpler or more involved than for a plurality contest. 
Suppose that the contest requires a candidate to receive at least a fraction $\phi \in (0, 1)$ of the valid votes
to be a winner.
(We allow the possibility that $\phi < 1/2$, in which case ``supermajority'' is a misnomer and there can be more than one winner;
this social choice function is used to determine `viability' in some U.S.\ partisan primaries.)
Suppose candidate $w$ is reported to be a winner.
Let $\theta_w$ denote the conditional probability that a ballot selected
at random from all ballot cards validly cast in the contest shows a valid vote for $w$, given that it shows a valid 
vote for any candidate in the contest,
and let $\eta_w$ be the number of votes reported for $w$, divided by the total valid votes reported in the contest.
BRAVO uses the SPRT to test the hypothesis $\theta_w = \phi$ against the alternative $\theta_w = \eta_w$
for each reported winner. 
If $\phi > 1/2$, there can be only one reported winner. 
If $\phi < 1/2$, there can be more than one, in which case
that hypothesis needs to be tested for all candidates (not just the reported winners), to confirm that (only) the
reported winner(s) won.
If it is reported that no candidate received at least a fraction $\phi$ of the valid votes,  
BRAVO tests the hypotheses that each candidate $\ell$  received $\phi$ of the valid votes
against the alternative that each candidate received $\eta_\ell < \phi$ of the valid votes,
to confirm that none received $\phi$ or more.

Consider independent, identically distributed (IID) draws from a binary population $\{x_i\}_{i=1}^N$, 
$x_i \in \{0, 1\}$ for all $i$.
Let $\theta = \bar{x} := \frac{1}{N} \sum_{i=1}^N x_i$ be the population fraction of 1s.
We sample with replacement from the population.
Let $X_k$ be the value selected on the $k$th draw. 
Then $\Pr \{X_k = 1 \} = \theta$ and $\Pr \{X_k = 0 \} = 1-\theta$.
By independence, the probability of a sequence $(X_k=y_k)_{k=1}^j$ is the product of the probabilities of the terms,
which can be written 
\begin{equation}
\Pr \left \{ \cap_{k=1}^j \{X_k = y_k \} \right \}= \prod_{k=1}^j \Pr \{X_k = y_k\} = \prod_{k=1}^j \left ( y_k \theta + (1-y_k)(1-\theta) \right ).
\end{equation}
The ratio of the probability of the sequence $(X_k=y_k)_{k=1}^j$ if $\theta = \eta$ to its probability if $\theta = \mu$ is
\begin{equation}
  \SPR_j = \prod_{k=1}^j \left ( y_k \frac{\eta}{\mu} + (1-y_k) \cdot \frac{1-\eta}{1-\mu} \right ).
\end{equation}
Wald's SPRT rejects the hypothesis that $\theta = \mu$ at significance level $\alpha$ if
$\SPR_j \ge 1/\alpha$ for any $j$.
That is, $\Pr_{\theta = \mu} \{\sup_j \SPR_j \ge 1/\alpha \} \le \alpha$: the SPRT is a \emph{sequentially valid} test.
Moreover, $\min(1, 1/\SPR_j)$ is an \emph{anytime $P$-value} for the hypothesis $\theta = \mu$; i.e.,
for any $p \in [0, 1]$, 
$$\Pr_{\theta = \mu}  \left \{\inf_{j=1}^\infty (1/\SPR_j) \le p \right \} \le p.$$
The SPRT is quite general; this is perhaps the simplest case.

Wald's proof that the general SPRT is sequentially valid is complicated, but Ville's inequality \citep{ville39} yields a simple proof.
Given a sequence of random variables $X_1, X_2, \ldots$, let $X^j$ denote the finite sequence $X_1, \ldots, X_j$.
A sequence of absolutely integrable random variables $T_1, T_2, \ldots$ is a \emph{martingale} with respect to a sequence of
random variables $X_1, X_2, \ldots$ if $\EE (T_j | X^{j-1}) = T_{j-1}$.
It is a \emph{supermartingale} if $\EE (T_j | X^{j-1}) \le T_{j-1}$.
The expected value of every term of a martingale is the same.
A (super)martingale is \emph{nonnegative} if $\Pr \{T_j \ge 0 \} = 1$ for all $j$.

Ville's inequality is an extension of Markov's inequality to supermartingales:
if $T_j$, $j=1, \ldots$, is a nonnegative supermartingale with respect to $X_j$, $j=1, \ldots$, then 
$$\Pr \{ \sup_{j \in \mathbb{N}} T_j \ge k\EE T_1\} \le 1/k.$$
The Bernoulli \SPR{} is a martingale with respect to $X_j$, $j=1, \ldots$, if $\theta = \mu$:
\begin{eqnarray}
\EE (\SPR_j | X^{j-1}) &=& \SPR_{j-1} \times \EE \left ( X_j \frac{\eta}{\mu} + (1-X_j) \frac{1-\eta}{1-\mu} \right ) \nonumber \\
&=& \SPR_{j-1} \times \left ( \mu \frac{\eta}{\mu} + (1-\mu) \frac{1-\eta}{1-\mu} \right )  \nonumber \\
&=& \SPR_{j-1} \times \left (\eta + (1-\eta) \right ) = \SPR_{j-1}.
\end{eqnarray}
Because $ \EE (\SPR_1) = 1$, Ville's inequality implies that $\Pr_{\theta = \mu} \{\sup_j \SPR_j \ge 1/\alpha \} \le \alpha$.
More generally, sequences of likelihood ratios are nonnegative martingales with respect to the distribution in the
denominator.

\citet{wald45} proved that among all sequentially valid tests of the hypothesis $\theta = \mu$, the SPRT with alternative
$\theta = \eta$ has the smallest expected sample size to reject $\theta=\mu$ when in fact $\theta = \eta$.
But when $\theta \in (\mu, \eta)$, the SPRT can fail to reject the null, continuing to sample forever,
and when $\theta > \eta$, the SPRT can be very inefficient.
As a result, when reported vote shares are incorrect but the reported winner(s) really won, BRAVO can require 
enormous samples, even when the true margin is large.

This paper introduces ALPHA, a simple adaptive extension of BRAVO.
It is motivated by the SPRT for the Bernoulli and its optimality when the simple alternative is true.
While BRAVO tests against the alternative that the true vote shares are equal to the reported vote shares,
ALPHA is adaptive, estimating the reported winner's share of the vote before the $j$th card is drawn
from the $j-1$ cards already in the sample.
The estimator can be any measurable function of the first $j-1$ draws that takes values in the composite alternative; 
numerical examples below use a simple truncated shrinkage estimate. 
ALPHA also generalizes BRAVO to situations where the population $\{x_j\}$ is not necessarily binary, 
but merely nonnegative and bounded.
That generalization allows ALPHA to be used with SHANGRLA to audit supermajority contests
and to conduct comparison audits of a wide variety of social choice functions---any for which there is a
SHANGRLA audit.
In contrast, BRAVO requires the list elements to be binary-valued.
Finally, ALPHA works for sampling with or without replacement, with or without weights, while BRAVO is specifically for 
IID sampling with replacement.
The SPRT for a population percentage using sampling without replacement is straightforward, but was not
in the original BRAVO paper \citep{lindemanEtal12}.

\section{ALPHA and SHANGRLA}

\subsection{SHANGRLA} \label{sec:shangrla}
Before introducing ALPHA, we provide additional motivation for constructing a more general test than BRAVO: 
the SHANGRLA framework for RLAs.
SHANGRLA \citep{stark20} checks outcomes by testing \emph{half-average assertions}, each of which claims 
that the mean of a finite list of
numbers between $0$ and $u$ is greater than $1/2$.
Each list of numbers results from applying an \emph{assorter} to the ballot cards.
The assorter uses the votes and possibly other information (e.g., how the voting system interpreted the ballot)
to assign a number between $0$ and $u$ to each ballot.
For some assorters, the numbers are only $0$ and $1$, but for others, there are more possible values.

The correctness of the outcomes under audit is implied by the intersection of a collection of such assertions;
the assertions depends on the social choice function, the number of candidates, and other details \citep{stark20}.
SHANGRLA tests the negation of each assertion, the
\emph{complementary null hypothesis} that each assorter mean is not greater than $1/2$.
If that hypothesis is rejected for every assertion,
the audit concludes that the outcome is correct.
Otherwise, the audit expands, potentially to a full hand count.
If every null is tested at level $\alpha$, this results in a risk-limiting
audit with risk limit $\alpha$: if the outcome is not correct, the chance the audit will stop
shy of a full hand count is at most $\alpha$.
No adjustment for multiple testing is needed \citep{stark20}.

The core, canonical statistical problem in SHANGRLA is to test the hypothesis that $\bar{x} \le 1/2$ using a 
sample from a finite population $\{x_i\}_{i=1}^N$, 
where each $x_i \in [0, u]$, with $u$ known.\footnote{%
An equivalent problem is to test the hypothesis that 
$\bar{y} \le t$ using a sample from $\{y_i\}_{i=1}^N$, where each $y_i \in [0, 1]$ (let $y_i = x_i/u$ and set $t=1/(2u)$).
}
This formulation unifies polling audits and comparison audits; the difference is only in how the 
values $\{x_i\}$ are calculated from the votes; see section~\ref{sec:comparison-audits}.
The sample might be drawn with or without replacement.
It might be drawn from the population as a whole 
(unstratified sampling), or the population might be divided into strata, each of which is sampled independently (stratified sampling).
It might be drawn using Bernoulli sampling, where each item is included independently, with some common probability.
Or batches of ballot cards might be sampled instead of individual cards (cluster sampling),
with equal or unequal probabilities; see section~\ref{sec:batch-audits}.

For instance, consider one reported winner and one reported loser in a single-winner or multi-winner plurality 
contest (any number of pairs can
be audited simultaneously using the same sample \citep{stark20}).
Let $N$ denote the number of ballot cards validly cast in the contest.
The assorter assigns the $i$th ballot the value $x_i=1$ if the ballot has a valid vote for the reported winner, 
the value $x_i=0$ if it has a valid vote for the reported loser, and the value $x_i=1/2$ otherwise.
The reported winner really beat the reported loser if $\theta := \frac{1}{N}\sum_i x_i > 1/2$.
In a multi-winner plurality contest with $W$ reported winners and $L$ reported losers,
the reported winners really won if the mean of each of the $WL$ lists for the (reported winner, reported loser) pairs
is greater than $1/2$.

\subsection{The ALPHA supermartingale test}

We start by developing a one-sided test of the simple hypothesis $\theta = \mu$, then show that the $P$-value is monotone in
$\mu$, so the test is valid for the composite hypothesis $\theta \le \mu$, as SHANGRLA requires.
Let $X^j := (X_1, \ldots, X_j)$.
Assume $X_i \in [0, u]$ for some known $u$. 
(For ballot-polling audits of plurality contests, $u=1$.)
Let $\mu_j := \EE(X_j | X^{j-1})$ computed under the null hypothesis $\theta = \mu$. 
Let $\eta_j = \eta_j(X^{j-1})$, $j=1, \ldots$, be a \emph{predictable sequence} in the sense that 
$\eta_j$ may depend on $X^{j-1}$, but not on $X_k$ for $k \ge j$.
We now define the ALPHA supermartingale $(T_j)_{j \in \mathbb{N}}$.
Let $T_0 := 1$ and
\begin{equation}
    T_j := T_{j-1} u^{-1}\left ( X_j\frac{\eta_j}{\mu_j} + (u-X_j) \frac{u-\eta_j}{u-\mu_j} \right ), \;\; j=1, \ldots . \label{eq:alphaBRAVOdef}
\end{equation}
This can be rearranged to yield
\begin{equation}
    T_j := T_{j-1} \left ( \frac{X_j}{\mu_j} \cdot \frac{\eta_j-\mu_j}{u-\mu_j} + \frac{u-\eta_j}{u-\mu_j} \right ). \label{eq:alphaMult}
\end{equation}
Equivalently,
\begin{equation}
    T_j := \prod_{i=1}^j \left ( \frac{X_i}{\mu_i} \cdot \frac{\eta_i-\mu_i}{u-\mu_i} + \frac{u-\eta_i}{u-\mu_i} \right ), \;\; j \ge 1. \label{eq:alphaMultProd}
\end{equation}
Under the null hypothesis that $\theta_j = \mu_j$, $T_j$ is nonnegative since $X_j$, $\mu_j$, and $\eta_j$
are all in $[0, u]$.
Also,
\begin{eqnarray}
    \EE (T_j | X^{j-1} ) &=& T_{j-1} \left ( \frac{\mu_j}{\mu_j} \cdot \frac{\eta_j-\mu_j}{u-\mu_j} + \frac{u-\eta_j}{u-\mu_j} \right ) \nonumber \\
    &=&  T_{j-1} \left ( \frac{\eta_j-\mu_j}{u-\mu_j} + \frac{u-\eta_j}{u-\mu_j} \right ) \nonumber \\
    &=& T_{j-1}.
\end{eqnarray}
Thus if $\theta = \mu$, $(T_j)_{j \in \mathbb{N}}$ is a nonnegative martingale with respect to $(X_j)_{j \in \mathbb{N}}$, starting at $1$.
If $\theta < \mu$, then $\EE (X_j | X^{j-1}) < \mu_j$ and $r_j = \frac{\EE (X_j | X^{j-1})}{\mu_j} < 1$, so
\begin{equation} \label{eq:supermartingale}
    \EE (T_j | X^{j-1} ) = T_{j-1} \left ( r_j \cdot \frac{\eta_j-\mu_j}{u-\mu_j} + \frac{u-\eta_j}{u-\mu_j} \right ) < T_{j-1}.
\end{equation}
Thus $(T_j)$ is a nonnegative supermartingale starting at 1 if $\theta \le \mu$.
It follows from Ville's inequality \citep{ville39} that if $\theta \le \mu$,
\begin{equation} \label{eq:p-value-adapt}
\Pr \{ \exists j :  T_j \ge \alpha^{-1} \} \le \alpha.
\end{equation}
That is, $\min(1, 1/T_j)$ is an ``anytime $P$-value'' for the composite null hypothesis $\theta \le \mu$.

Note that the derivation did not use any information about $\{x_i\}$ other than $x_i \in [0, u]$:
it applies to populations $\{x_i\}$ that are nonnegative and bounded, not merely binary populations.
Hence, it can be used to test \emph{any} SHANGRLA assertion, including those for
a wide variety of social choice functions---plurality, multi-winner plurality, super-majority, d'Hondt and other
proportional representation schemes, Borda count, approval voting, STAR-Voting, arbitrary scoring rules, and 
IRV---using sampling with or without replacement, with or without stratification.
The ALPHA supermartingales comprise the same family of betting supermartingales studied by 
\citet{waudby-smithRamdas21,waudby-smithEtal21}, but are parametrized
differently; see section~\ref{sec:rilacs} below.

\subsubsection{Sampling without replacement}
To use ALPHA with a sample drawn without replacement, we need $\EE(X_j | X^{j-1})$ computed on the assumption that
$\theta :=  \frac{1}{N} \sum_{i=1}^N x_i = \mu$.
For sampling without replacement from a population with mean $\mu$, after draw $j-1$, the mean of the remaining numbers is $(N\mu - \sum_{k=1}^{j-1}X_k)/(N-j+1)$.
Thus the conditional expectation of $X_j$ given $X^{j-1}$ under the null is $(N\mu - \sum_{k=1}^{j-1}X_k)/(N-j+1)$.
If $N\mu - \sum_{k=1}^{j-1}X_k < 0$ for any $k$, the null hypothesis $\theta = \mu$ is certainly false.

\subsubsection{BRAVO is a special case of ALPHA}
BRAVO is ALPHA with the following restrictions:
\begin{itemize}
    \item the sample is drawn with replacement from ballot cards that do have a valid vote for the reported winner 
    $w$ or the reported loser $\ell$ (ballot cards with votes for other candidates or non-votes are ignored)
    \item  ballot cards are encoded as 0 or 1, depending on whether they have a valid vote
    for the reported winner or for the reported loser;  $u=1$ and the only possible values of $x_i$ are 0 and 1
    \item $\mu = 1/2$, and $\mu_i = 1/2$ for all $i$ since the sample is drawn with replacement
    \item  $\eta_i = \eta_0 := N_w/(N_w+N_\ell)$, where $N_w$ is the number of votes reported for candidate $w$ 
and $N_\ell$ is the number of votes reported for candidate $\ell$: $\eta$ is not updated as data are collected
\end{itemize}
It follows from \citet{wald45} that BRAVO minimizes the expected 
sample size to reject the null hypothesis $\theta=1/2$ when $w$ really received
the share $\eta_0$ of the reported votes.
The motivation for this paper is that $w$ almost never receives \emph{exactly} their reported vote share, and
BRAVO (and other RLA methods that rely on the reported vote share) may then have poor performance---even
though they are still guaranteed to limit the risk that the audit will not correct an incorrect result to at most $\alpha$.

When the reported vote shares are incorrect, using a method that
adapts to the observed audit data can help, as we shall see.

\subsection{Relationship to RiLACS and Betting Martingales} \label{sec:rilacs}
\citet{waudby-smithRamdas21,waudby-smithEtal21} develop tests and
confidence sequences for the mean of a bounded population
using \emph{betting martingales} of the form
\begin{equation} \label{eq:lambda-rilacs}
M_j := \prod_{i=1}^j (1 + \lambda_i (X_i- \mu_i)),
\end{equation}
where, as above, $\mu_i := \EE(X_i | X_{i-1})$, computed on the assumption that the null hypothesis is true.
The sequence $(M_j)$ can be viewed as the fortune of a gambler in a series of wagers.
The gambler starts with a stake of $1$ unit and bets a fraction $\lambda_i$ of their current wealth on the outcome
of the $i$th wager.
The value $M_j$ is the gambler's wealth after the $j$th wager.
The gambler is not permitted to borrow money, so to ensure that when $X_i = 0$ (corresponding to losing the $i$th bet)
the gambler does not end up in debt ($M_i < 0$), $\lambda_i$ cannot exceed $1/\mu_i$.

The ALPHA supermartingale is of the same form:
\begin{eqnarray} \label{eq:lambda-form}
T_j &=& \prod_{i=1}^j \left ( \frac{X_i}{\mu_i} \cdot \frac{\eta_i-\mu_i}{u-\mu_i} + \frac{u-\eta_i}{u-\mu_i} \right ) \nonumber \\
&=& \prod_{i=1}^j \frac{X_i (\eta_i/\mu_i -1) + u - \eta_i}{u-\mu_i} \nonumber \\
&=& \prod_{i=1}^j \left ( 1 + \frac{X_i (\eta_i/\mu_i -1) + \mu_i - \eta_i}{u-\mu_i} \right ) \nonumber \\
&=&  \prod_{i=1}^j \left ( 1 + \frac{\eta_i/\mu_i -1}{u-\mu_i} \cdot (X_i - \mu_i) \right ),
\end{eqnarray}
identifying $\lambda_i \equiv \frac{\eta_i/\mu_i -1}{u-\mu_i}$.
Choosing $\lambda_i$ is equivalent to choosing $\eta_i$:
\begin{equation}
\lambda_i = \frac{\eta_i/\mu_i -1}{u-\mu_i} \;\; \Longleftrightarrow \;\; \eta_i = \mu_i \left ( 1 + \lambda_i (u-\mu_i) \right ).
\end{equation}
As $\eta_i$ ranges from $\mu_i$ to $u$, $\frac{\eta_i/\mu_i -1}{u-\mu_i}$ ranges continuously from
0 to $1/\mu_i$, the same range of values of $\lambda_i$ permitted in \citet{waudby-smithRamdas21,waudby-smithEtal21}:
selecting $\lambda_i$ is equivalent to selecting a method for estimating $\theta_i$.
That is, the ALPHA supermartingales are identical to the betting martingales in \citet{waudby-smithRamdas21,waudby-smithEtal21};
the difference is only in how $\lambda_i$ is chosen.
(However, see section~\ref{sec:batch-audits} for a generalization to allow sampling weights and to allow $u$ to
vary by draw.)

\citet{waudby-smithRamdas21,waudby-smithEtal21} consider two classes of strategies for picking $\lambda_i$ 
intended to maximize the expected rate at which the gambler's wealth grows. 
One of the classes is approximately optimal if $\theta$ is known (much like BRAVO is optimal when
the reported results are correct); the other does not user prior information,
instead using the data to adapt to the true value of $\theta$.
The ALPHA representation of the betting martingales provides a family of tradeoffs between those 
extremes, using different estimates
of $\theta_i$ based on $\eta$ and $X^{i-1}$.
Parametrizing the selection of $\lambda_i$ in terms of an estimate $\eta_i$ of $\theta_i$ 
may aid intuition in developing
more powerful supermartingale tests (in the sense that they tend to reject sooner) 
for particular applications---such as election audits.

\subsection{Comparison audits} \label{sec:comparison-audits}
In the SHANGRLA framework, there is no formal difference between \emph{polling audits} (which do not use the voting system's
interpretation of ballot cards) and \emph{comparison audits}, which involve comparing how the voting system interpreted cards to 
how humans interpret the same cards.
Either way, the correctness of the election outcome is implied by a collection of assertions, each of which is of the form, 
``the average of this list of $N$ numbers in $[0, u]$ is greater than $1/2$.''
The only difference is the particular function that assigns numbers in $[0, u]$ to ballot cards.
For polling audits, the number assigned to a card depends on the votes on that card as interpreted by a human (and on the social
choice function and other parameters of the contest), but not on how the voting system interpreted the card.
For comparison audits, the number also depends on how the system interpreted that card and on the reported ``assorter margin.''
See \citet[Section 3.2]{stark20} for details.

Because ALPHA can test the hypothesis that the mean of a bounded, nonnegative population is not greater than 1/2 
(even for populations with more than two values), it works for comparison and polling audits with no modification.
The interpretation of $\theta$ is different: instead of being related to vote shares, it is related to the amount of \emph{overstatement error}
in the system's interpretation of each ballot card.
For comparison audits, the initial value for the alternative, $\eta_0$, could be chosen by making assumptions 
about how often the system made errors of various kinds.
The risk is rigorously limited even if those assumptions are wrong, but the choice affects the performance.

To conduct a comparison RLA, auditors export subtotals or other vote records from the voting system and \emph{commit to them} 
(e.g., by publishing them).
Election auditors first check whether applying the social choice function to the exported records gives the same outcome
reported for each contest.
If not, the election fails the audit: even according to the voting system, some
reported outcome is wrong.
If the reported outcomes match those implied by the exported vote records, the audit next checks whether differences between the voting system's exported records and a human interpretation
of the votes on ballot cards could have altered any reported outcome, by
manually checking a random sample of the voting system's exported records against a manual interpretation of
the votes on the corresponding physical batches of ballot cards.

This procedure is like checking an expense report. 
Committing to the subtotals is like submitting the expense report.
An auditor can check the accuracy of the report by first checking the addition 
(checking whether the exported batch-level results produce the reported contest outcomes), then manually checking a sample of 
the reported expenses against the physical paper receipts (checking the accuracy of the machine interpretation of the cards).

\subsection{Setting $\eta_i$ to be an estimate of $\theta_i$}
Since the SPRT minimizes the expected sample size to reject the null when the alternative is true,
we might be able to construct an efficient test by using as the alternative an estimate of $\theta$
based on the audit data and the reported results.
Any estimate $\eta_i$ of $\theta_i$ that does not depend on $X_k$ for $k \ge i$ preserves the supermartingale property
under the null, 
and the auditor has the freedom to ``change horses'' and use a different estimator
at will as the sample evolves. 
For example, $\eta_i$ might be constant, as it is in BRAVO.
Or it could be constant for the first 100 draws, then switch to the unbiased estimate of $\theta_i$ 
based on $X^{i-1}$ once $i \ge 100$.
Or it could be a Bayes estimate of $\theta_i$ using data $X^{i-1}$ and a prior concentrated on
$[\mu_0, u]$, centered at the value of $\theta$ implied by the reported results.
(The $P$-value of the test is still a frequentist $P$-value; the estimate $\eta_i$ affects the power.)
Or it could give the sample mean a weight that grows as the sample standard deviation shrinks.
Or it could be the estimate implied by choosing $\lambda_i$ using one of the methods for selecting $\lambda_i$.
described by \citet{waudby-smithRamdas21}.

\subsubsection{Naively maximizing $\EE (T_i| X^{i-1})$ does not work}
Suppose that $\theta_i := \EE (X_i | X^{i-1}) > \mu_i$, i.e., that the alternative hypothesis is true. 
What value of $\eta_i$ maximizes $\EE (T_i | X_{i-1})$?
\begin{eqnarray}
\EE \left ( \left . \frac{X_i}{\mu_i} \cdot \frac{\eta_i-\mu_i}{u-\mu_i} + \frac{u-\eta_i}{u-\mu_i} \right | X^{j-1} \right ) &=&
\frac{\theta_i}{\mu_i} \cdot \frac{\eta_i-\mu_i}{u-\mu_i} + \frac{u-\eta_i}{u-\mu_i} \nonumber \\
&=& \eta_i \left ( \frac{\frac{\theta_i}{\mu_i} - 1}{u-\mu_i} \right ) + \frac{u-\theta_i}{u-\mu_i}.
\end{eqnarray}
This is monotone increasing in $\eta_i$, so it is maximized for $\eta_i = u$, for a single draw.
But if $\eta_i=u$ and $X_i = 0$, then $T_j = 0$ for all $j \ge i$, and the test will never reject the null
hypothesis, no matter how many more data are collected.
This is essentially the observation made by \citet{kelly56}, leading to the
Kelly criterion.
Keeping $\eta_i < u$ hedges against that possibility.

Instead of picking $\eta_i$ to maximize the next term $T_i$, one can pick it to maximize the rate at which $T$ grows.
In the binary data case, the Kelly criterion \citep{kelly56}, discussed by \citet{waudby-smithRamdas21}, leads to the
optimal choice when $\theta$ is known.
For sampling with replacement, this is $\lambda = 2(N_w-N_\ell)/(N_w+N_\ell) = 4(\theta-1/2)$, since $N_w/(N_w+N_\ell) = \theta$.
This corresponds to $\eta =  (1/2) \left ( 1 + 4(\theta-1/2)  (1-1/2) \right ) = \theta$, the population mean. 

\subsubsection{Illustration: a simple way to select $\eta_i$}
Any choice of $\eta_i \in (\mu_i, u)$ that depends only on $X^{i-1}$ preserves the supermartingale property
under the composite null, and thus
the validity of the ALPHA test.
To show the potential of ALPHA, the simulations reported below are based on setting $\eta_i$ to be a simple 
``truncated shrinkage'' estimate of $\theta_i$.
The estimator shrinks towards the reported result as if the reported
result were the mean of 
$d>0$ draws from the population ($d$ is not necessarily an integer). 
To ensure that the alternative hypothesis corresponds to the reported
winner really winning, we need $\eta_i > \mu_i$, and to keep the estimate consistent with the
constraint that $x_i \in [0, u]$, we need $\eta_i \le u$.
The following estimate $\eta_i$ of $\theta_i$ meets both requirements:
\begin{equation} \label{eq:etaDef}
\eta_i :=  \left ( \frac{d\eta_0 + \sum_{k=1}^{i-1}X_i }{d+i-1} \vee (\mu_i+\epsilon_i ) \right )
\wedge u.
\end{equation}

{\bf Choosing $\eta_0$.}
The starting value $\eta_0$ could be the value of $\theta$ implied by the reported results.
For a polling audit, that might be based on the reported margin in a plurality contest.
For a comparison audit, that might be based on historical experience with tabulation error.
But the procedure could be made fully adaptive by starting with,
say, $\eta_0 = (u+\mu)/2$ or $\eta_0 = u$.

{\bf Choosing $d$.}
As $d \rightarrow \infty$, the sample size for ALPHA approaches that of BRAVO, for binary data.
The larger $d$ is, the more strongly anchored the estimate is to the reported vote shares, and
the smaller the penalty ALPHA pays when the reported results are exactly correct.
Using a small value of $d$ is particularly helpful when the true population mean is far from the reported results.
The smaller $d$ is, the faster the method adapts to the true population mean, but the higher the variance is.
Whatever $d$ is, the relative weight of the reported vote shares decreases as the sample size increases.

{\bf Choosing $\epsilon_i$.}
To allow the estimated winner's share $\eta_i$ to approach
$\mu_i$ as the sample grows (if the sample mean approaches $\mu_i$ or less),
we shall take $\epsilon_i := c/\sqrt{d+i-1}$ for a nonnegative constant $c$,
for instance $c=(\eta_0-\mu)/2$.
The estimate $\eta_i$ is thus the sample mean,
shrunk towards $\eta_0$ and truncated to the interval $[\mu_i + \epsilon_i, 1)$, where
$\epsilon_i \rightarrow 0$ as the sample size grows.

\section{Pseudo-algorithm for ballot-level comparison and ballot-polling audits}

The algorithm below is written for a single SHANGRLA assertion, but the audit can be conducted in parallel
for any number of assertions using the same sampled ballot cards; no multiplicity adjustment for the number
of assertions is needed.
There are assorters for polling audits, which do not use information about how the voting system interpreted
ballot cards, and for comparison audits, which require the voting system to commit to how it interpreted
each ballot card before the audit starts.
For comparison audits, the first step is to verify that the data exported from the voting system
reproduces the reported election outcome, that is, to check whether applying the social choice function to the
cast vote records gives the same winners.
We shall assume that a compliance audit has shown that the paper trail is trustworthy.
For comparison audits, we assume that the system has exported a CVR for every ballot card.
(For methods to deal with a mismatch between the number of ballot cards and the number of CVRs, see
\citet{stark20}.)

\begin{itemize}
   \item Set audit parameters:
       \begin{itemize}
          \item Select the risk limit $\alpha \in (0, 1)$; decide whether to sample with or without replacement.
          \item Set $u$ as appropriate for the assertion under audit.
          \item Set $N$ to the number of ballot cards in the population of cards from which the sample is drawn.
          \item Set $\eta_0$. For polling audits, $\eta_0$ could be the reported mean value of the assorter. (For instance, for the assertion corresponding
          to checking whether $w$ got more votes than $\ell$,  $\eta_0 = (N_w + N_c/2)/N$, where $N_w$ is the number of
          votes reported for $w$, $N_\ell$ is the number of votes reported for $\ell$, and $N_c = N-N_w-N_\ell$ is the number
          of ballot cards reported to have a vote for some other candidate or no valid vote in the contest.)
          For comparison audits, $\eta_0$ can be based on assumed or historical rates of overstatement errors.
          \item Define the function to update $\eta$ based on the sample, e.g., \\
          $\eta(i, X^{i-1}) = \left ( (d\eta_0 + S)/(d+i-1) \vee (\epsilon(i)+ \mu_i) \right ) \wedge u$, where
          $S = \sum_{k=1}^{i-1}X_k$ is the sample sum of the first $i-1$ draws and $\epsilon(i) = c/\sqrt{d+i-1}$;
          set any free parameters in the function (e.g., $d$ and $c$ in this example).
          The only requirement is that $\eta(i, X^{i-1}) \in (\mu_i, u)$, where $\mu_i := \EE (X_i | X^{i-1})$ is computed under the null.
      \end{itemize}
    \item Initialize variables
        \begin{itemize}
          \item $j \leftarrow 0$: sample number
          \item $T \leftarrow 1$: test statistic
          \item $S \leftarrow 0$: sample sum
          \item $m = 1/2$: population mean under the null
      \end{itemize}
   \item While $T < 1/\alpha$ and not all ballot cards have been audited:
   \begin{itemize}
        \item Draw a ballot at random
        \item $j \leftarrow j+1$
        \item Determine $X_j$ by applying the assorter to the selected ballot card (and the CVR, for comparison audits)
        \item If $m < 0$, $T \leftarrow \infty$. Otherwise, $T \leftarrow T u^{-1} \left ( X_j\frac{\eta(j, S)}{m} + (u-X_j) \frac{u-\eta(j,S)}{u-m} \right )$; 
        \item $S \leftarrow S+X_j$
        \item If the sample is drawn without replacement, $m \leftarrow (N/2 - S)/(N-j+1)$
        \item If desired, break and conduct a full hand count instead of continuing to audit. 
     \end{itemize}
     \item If a full hand count is conducted, its results replace the reported results if they differ.
\end{itemize}

\section{Batch-Polling and Batch-Level Comparison Audits} \label{sec:batch-audits}

So far we have been discussing audits that sample and manually interpret individual ballot cards: \emph{ballot-polling} audits,
which use only the manual interpretation of the sampled ballots, and \emph{ballot-level comparison} audits, which also use the system's
interpretation of the sampled ballot cards (CVRs).
Ballot-level comparison audits are the most efficient strategy (measured by expected sample size) if the voting system can export CVRs in a
way that the allows the corresponding physical ballot cards to be identified, retrieved, and interpreted manually.
Legacy voting systems cannot: some do not create CVRs at all, and some that do create CVRs do not provide information to link
each CVR to the corresponding physical card.
Even with modern equipment, reporting CVRs linked to physical ballot cards while maintaining vote anonymity is hard
if votes are tabulated in precincts or vote centers, because the order in which ballot cards are scanned, tabulated, and stored 
can be nearly identical to the order in which they were cast. 
(However, see \cite{stark22b}.)

Many jurisdictions tabulate and store ballot cards in physical batches for which the voting system
can report batch-level results.\footnote{%
Vote centers and vote-by-mail can make batch-level comparison audits hard or impossible,
since some voting systems can only report vote subtotals for batches based on political geography (e.g., precincts),
which may not correspond to physically identifiable batches.
To create physical batches that match the reporting batches would require sorting the ballot cards.
}
Thus it can be desirable to sample and interpret \emph{batches} of ballot cards instead of \emph{individual} ballot cards.
\emph{Batch-polling} audits use human interpretation of the votes in the batches but not the voting system's tabulation (other than the system's report of who won, and possibly the reported vote totals).
\emph{Batch-level comparison} audits compare human interpretation of the ballot cards in the sampled batches to the voting system's
interpretation of the same cards.
Batch-level comparison audits are operationally similar to existing audits in many states,
including California and New York---but RLAs provide statistical guarantees that those statutory audits do not
provide.

For many social choice functions (including all scoring rules), knowing the total number of votes
reported for each candidate in each batch is enough to conduct a batch-level comparison audit.
But for some voting systems and some social choice functions, batch-level results contain too little information.
For instance, to audit instant-runoff voting (IRV), it is not enough to know how many voters gave each rank to each candidate:
the joint distribution of ranks matters.

As discussed above, SHANGRLA audits of one or more contests involve a collection of assorters $\{A_j\}_{j=1}^A$,
functions from ballot cards (and possibly additional information, such as the reported outcome, reported margin,
and the system's interpretation of the votes on the ballot card) to $[0, u_j]$.
The domain of assorter $j$ is $\cD_j$, which could comprise all ballot cards cast in the election
or a smaller set, provided $\cD_j$ includes every card
that contains the contest that assorter $A_j$ is relevant for.
Targeting audit sampling using information about which ballot cards purport to contain which contests (\emph{card style} data)
can vastly improve audit efficiency while rigorously maintaining the risk limit even if the voting system misidentifies which
cards contain which contests \citep{glazerEtal21}.
There are also techniques for dealing with missing ballot cards \citep{banuelosStark12,stark20}.

Let $|\cD_j|$ denote the number of ballot cards in $\cD_j$.
Every audited contest outcome is correct if every assorter mean is greater than $1/2$, i.e., if for all $j$,
\begin{equation}
    \bar{A}_j := \frac{1}{|\cD_j|} \sum_{b_i \in \cD_j } A(b_i) > 1/2.
\end{equation}
Ballot cards cards are tabulated and stored in disjoint \emph{batches}
$\{ \cB_k \}$ of physically identifiable ballot cards.
Let $|\cB_k|$ be the number of ballot cards in batch $k$.
We assume that each assorter domain $\cD_j$ is the union of some of the batches:
$\cD_j = \cup_{k: \cB_k \subset \cD_j} \cB_k$.
Let $\mc{K}_j  := \{k: \cB_k \subset \cD_j\}$ be the indices of the batches to which assorter $A_j$ applies
and let $|\mc{K}_j|$ denote the cardinality of $\mc{K}_j$.
Define
\begin{equation}
   A_{jk} := \sum_{b_i \in \cB_k} A_j(b_i),
\end{equation}
the total of assorter $j$ over batch $k$.
Then $ \bar{A}_j = \frac{1}{|\cD_j|} \sum_{k \in \mc{K}_j} A_{jk}$.
Let $u_{jk}$ be an upper bound on $A_{jk}$, for instance $u_j |\cB_k|$.
Tighter upper bounds than that may be calculable, in particular for batch comparison audits:
depending on the reported votes in batch $\cB_k$, the upper bound $u_j$ might not be attainable for every ballot card
in the batch.
Let $U_j := \sum_{k \in \mc{K}_j} u_{jk}$ be the sum of the batch upper bounds.

\subsection{Batch Sampling with Equal Probabilities}

Define
\begin{equation}
    \widetilde{A}_{jk} := A_{jk} \cdot \frac{|\mc{K}_j|}{|\cD_j|}.
\end{equation}
Then
\begin{equation}
    \frac{1}{|\mc{K}_j|} \sum_{k \in \mc{K}_j} \widetilde{A}_{jk} =  \bar{A}_j.
\end{equation}
That is, the mean of the $|\mc{K}_j|$ values $\{ \widetilde{A}_{jk}\}_{k \in \mc{K}_j}$ is equal to
the mean of the $|\cD_j|$ values $\{A_j(b_i)\}_{b_i \in \cD_j}$.
Let $\widetilde{u}_{jk} = u_{jk} \frac{|\mc{K}_j|}{|\cD_j|}$
and $\widetilde{u}_j := \max_{k \in \mc{K}} \widetilde{u}_{jk}$.
Then $\{ \widetilde{A}_{jk} \}_{k \in \mc{K}_j}$ are in $[0, \widetilde{u}_j]$, so
if we sample batches with equal probability (with or without replacement), testing whether the population mean $\bar{A}_{j}$
is less than or equal to $1/2$ is an instance of the problem solved by ALPHA,
the tests in \citet{waudby-smithRamdas21,waudby-smithEtal21}, and the Kaplan martingale test;
for sampling with replacement, it is also solved by the Kaplan-Wald and Kaplan-Markov tests.

However, because batch sizes may vary widely, using a single upper bound $\widetilde{u}_j$ for all batches may have a great 
deal of slack for some batches, which can reduce power.
By sampling batches with unequal probabilities, we can transform the problem to one where the upper bounds on the
batches are sharper. 
This may lead to more efficient audits, depending on fixed costs related to retrieving batches; checking, recording, and opening seals; 
re-sealing batches and returning them to storage; etc.

\subsection{Batch Sampling with Probability Proportional to a Bound on the Assorter}

Let $K_i$ denote the batch selected in the $i$th draw.
For sampling without replacement, let $\mc{K}_{j\ell} = \mc{K}_j \setminus \{K_i\}_{i=1}^{\ell-1}$; for sampling
with replacement, let $\mc{K}_{j\ell} = \mc{K}_j$.
For sampling without replacement, let $\cD_{j\ell} = \cup_{k \in \mc{K}_{j\ell}} \cB_k$;
for sampling with replacement, let $\cD_{j\ell} = \cD_j$.
Then $\mc{K}_{j\ell}$ are the indices of the batches from which the
$\ell$th sample batch will be drawn, and $\cD_{j\ell}$ are the ballot cards those batches contain.
Let $U_{j\ell} := \sum_{i \in \mc{K}_{j\ell}} u_{ji}$.
The $\ell$th batch is selected at random from $\{\cB_k : k \in \mc{K}_{j\ell} \}$, with chance $u_{jk}/U_{j\ell}$ 
of selecting $\cB_k$.

Define
\begin{equation}
    \hat{A}_{jk\ell} := A_{jk} \frac{U_{j\ell}}{u_{jk}|\cD_{j\ell}|} \in [0, \hat{u}_{j\ell}],
\end{equation}
where $\hat{u}_{j\ell} := U_{j\ell}/|\cD_{j\ell}|$.
(For sampling without replacement, this typically varies with $\ell$.)
Let $X_i := \hat{A}_{jK_ii}$ be the value of $\{ \hat{A}_{jki}\}_{k \in \mc{K}_{jk}}$ selected on the $i$th draw.
Consider the expected value of $X_i$ given $X^{i-1}$:
\begin{eqnarray}
    \theta_{ji} := \mathbb{E} (X_i | X^{i-1} ) 
    &=&  \sum_{k \in \mc{K}_{ji}} \frac{u_{jk}}{U_{ji}} A_{jk} \frac{U_{ji}}{u_{jk}|\cD_{ji}|} \nonumber \\
    &=& \frac{1}{|\cD_{ji}|} \sum_{k \in \mc{K}_{ji}} A_{jk},
\end{eqnarray}
the mean value of the assorter $A_j$ over the ballots that remain in the population just before the $i$th draw.
Under the null hypothesis that $\theta_j := \bar{A}_j \le \mu_j$,
\begin{eqnarray}
  \theta_{ji} \le \frac{|\cD_j| \mu_j - \sum_{k=1}^{i-1} A_{jK_k}}{|\cD_{ji}|} =: \mu_{ji}.
\end{eqnarray}
Let $\eta_{ji} \in (\mu_{ji}, \hat{u}_{ji}]$ be an estimate of $\theta_{ji}$ based on $X^{i-1}$ and define
\begin{equation}
    T_{jk} := \prod_{i=1}^k \left ( \frac{X_i}{\mu_{ji}} \cdot \frac{\eta_{ji}-\mu_{ji}}{\hat{u}_{ji}-\mu_{ji}} +
     \frac{\hat{u}_{ji}-\eta_{ji}}{\hat{u}_{ji}-\mu_{ji}} \right ).
\end{equation}
This generalizes ALPHA by allowing the population upper bound $\hat{u}_{ji}$ to vary from draw to draw, with a corresponding
draw-dependent constraint on $\eta_{ji}$.
As before, under the null hypothesis that $\theta_j \le \mu_j$,
$\{T_i\}$ is a nonnegative supermartingale starting at $1$:
$\eta_{ji} > \mu_{ji}$,
$\EE (X_i | X^{i-1}) \le \mu_{ji}$, and $r_i := \EE (X_i | X^{i-1})/\mu_{ji} \le 1$, so
\begin{equation} \label{eq:supermartingale-batch}
    \EE (T_i | X^{i-1} ) = T_{i-1} \left ( r_i \cdot \frac{\eta_{ji}-\mu_{ji}}{u_{ji}-\mu_{ji}} + \frac{u_{ji}-\eta_{ji}}{u_{ji}-\mu_{ji}} \right ) \le T_{i-1}.
\end{equation}
Thus by Ville's inequality, if $\theta_j \le \mu_j$,
\begin{equation}
\Pr_{\theta_j \le \mu_j} \{ \max_k T_{jk} \ge \alpha^{-1} \} \le \alpha.
\end{equation}

\subsubsection{Auditing many assertions using the same weighted sample of batches}
To audit more than one assertion using the same sample of batches, the sampling weights,
and thus the batch-level \emph{a priori} bounds, need to be commensurable:
if batches $\cB_\ell$ and $\cB_m$ are relevant for assorters $A_j$ and $A_k$, then we need
$u_{j\ell}/u_{k \ell} = u_{jm}/u_{km}$.
The easiest way to accomplish that is to take $u_{jm} = u_j |\cB_m|$ for $j = 1, \ldots, A$ and $m \in \cD_j$.
Tighter bounds may be possible in some cases, depending on the batch-level reports for all the contests
under audit.

\section{Stratified Sampling} \label{sec:stratified}
Stratified sampling---partitioning ballot cards into disjoint strata and sampling independently from those 
strata---can be helpful in RLAs  \citep{stark08a,higginsEtal11,ottoboniEtal18,stark20}.
For instance, some states (including California) require jurisdictions to draw audit samples independently.
Auditing a cross-jurisdictional contest then involves stratified samples; each stratum consists of the ballot
cards cast in one jurisdiction.
Stratified sampling can also offer logistical advantages by making it possible to use different audit strategies 
(polling, batch polling, ballot-level comparison, batch-level comparison) for 
different subsets of ballot cards, for instance, if some ballot cards are tabulated using equipment that can report how it
interpreted each ballot and some are not.

Stratified batch-comparison RLAs were developed in the first paper on risk-limiting audits, \citet{stark08a}.
The approach was tightened in \citet{higginsEtal11}.
\citet{ottoboniEtal18} developed a more flexible approach, SUITE (Stratified Union-Intersection Tests of Elections),
which does not require using the same sampling or audit strategy in different strata.
In particular, SUITE allows using polling in some strata and ballot-level or batch-level comparisons in others.
BRAVO does not work for auditing in the polling strata in that context, because it makes inferences about the votes 
for one candidate as a fraction of the votes that are either for that candidate or one other candidate, that is, it
conditions on the event that the selected card has a vote for either the reported winner or the reported loser.
That suffices to tell who won a plurality contest---by auditing every (reported winner, reported loser) 
pair---if all the ballot cards are in a single stratum, but not when the sample is stratified.

When the sample is stratified, what is needed is an inference about the \emph{number} of votes in the stratum for each candidate.
To solve that problem, \citet{ottoboniEtal18} used a test in the polling stratum based on the multinomial distribution, 
maximizing the $P$-value over a nuisance parameter, the number of ballot cards in the stratum with no valid vote for either candidate.
SUITE represents the hypothesis that the outcome is wrong as a union of intersections of hypotheses.
The union is over all ways of partitioning outcome-changing errors across strata.
The intersection is across strata for each partition in the union.
For each partition, for each stratum, SUITE computes a $P$-value for the hypothesis that the error in that stratum
exceeds its allocation, then combines those $P$-values across strata (using a combining
function such as Fisher's combining function) to test the intersection hypothesis that the error in 
every stratum exceeds its allocation in the partition.
If the maximum $P$-value of that intersection hypothesis over all allocations of outcome-changing error is less
than or equal to the risk limit, the audit stops. 
\citet{stark20} extends the union-intersection approach to use SHANGRLA assorters,
avoiding the need to maximize $P$-values over nuisance parameters in individual strata and
permitting sampling with or without replacement.

\subsection{ALPHA obviates the need to use a combining function across strata}
Because ALPHA works with polling and comparison strategies, it can be the basis of the test in every stratum,
whereas SUITE used completely different ``risk measuring functions'' for strata where the audit involves ballot polling and strata
where the audit involves comparisons.
We shall see that this obviates the need to use a combining function to combine $P$-values across
strata: the test supermartingales can just be multiplied, and the combined $P$-value is the reciprocal of their product.
This is because (predictably) multiplying terms in the product representation of different sequences---each of which, under the nulls
in the intersection, is a nonnegative supermartingale
starting at one---yields a nonnegative supermartingale starting at one.
Thus the product of the stratum-wise test statistics in any order (including interleaving terms across strata) is also a test statistic 
with the property that the chance it is greater than or equal to $1/\alpha$ is at most $\alpha$ under the intersection null.
Because Fisher's combining function adds two degrees of freedom to the chi-square distribution for each stratum, avoiding 
the need for a combining function can substantially increase power as the number of strata grows.
Table~\ref{tab:fisher} illustrates this increase: it shows the combined $P$-value for the intersection hypothesis when the 
$P$-value in each
stratum is $0.5$. 
The number of strata ranges from $2$---which might arise in an audit in a single jurisdiction when
stratifying on mode of voting (in-person versus absentee)---to 150---which might arise in auditing a cross-jurisdictional
contest in a state with many counties.
For instance, Georgia has 159 counties, Kentucky has 120, Texas has 254, and Virginia has 133.

\begin{table} 
\centering
\begin{tabular}{r|cc} 
 strata & Fisher's combination & supermartingale $P$ \\
 \hline
2 & 0.5966 & 0.25000000 \\
5 & 0.7319 & 0.03125000 \\
10 & 0.8374 & 0.00097656 \\
25 & 0.9514 & 0.00000003 \\
50 & 0.9917 & 0.00000000 \\
100 & 0.9997 & 0.00000000 \\
150 & 1.0000 & 0.00000000 
\end{tabular} 
 \caption{\protect \label{tab:fisher} Overall $P$-value for the intersection null hypothesis if the $P$-value in each stratum is 0.5, for Fisher's combining function (column~2) and for supermartingale-based tests (column~3). 
 The ``stratification penalty'' arising from the large number of degrees of freedom (twice the number of strata) for Fisher's combining function can be avoided by using supermartingale-based tests, which permit simply multiplying the test statistics across strata and taking
 the reciprocal of the result  (or 1, if 1 is smaller) as the $P$-value.}
 \end{table}

\subsection{Supermartingale-based tests of intersection hypotheses} \label{sec:supermartingale-intersection}
Here is a sketch of how ALPHA can be used for stratified audits.
Suppose there are $N$ ballots in all, partitioned into $S$ strata.
(This section will overload $S$ to mean two related things: the number of strata and
a mapping $S(\cdot)$ from counting numbers to strata. 
Elsewhere in the paper, $S_j$ refers to a sample sum.)
Stratum $s$ contains $N_s$ ballot cards; $\sum_s N_s = N$.
We want to test the hypothesis that $\bar{A} \le 1/2$.
Let $u$ be the upper bound on the numbers $A$ assigns.
Let $\bar{A}_s$ be the average of the assorter restricted to stratum $s$, so
$\bar{A} = N^{-1}\sum_s N_s \bar{A}_s$.
Suppose $\boldsymbol{\mu} = (\mu_s)_{s=1}^S$ satisfies $0 \le \mu_s \le u$.
We sample independently from the strata.
Let $X_{si}$ denote the $i$th draw from the $s$th stratum, and define $\mu_{si}$, $u_{si}$, and $\eta_{si}$ analogously.
Define
\begin{equation}
   R_{si}(\mu_s) := \frac{X_{si}}{\mu_{si}} \cdot \frac{\eta_{si}-\mu_{si}}{u-\mu_{si}} + \frac{u-\eta_{si}}{u-\mu_{si}}.
 \end{equation}
Recall from equation~\ref{eq:alphaMultProd} that
\begin{equation} 
T_j^s(\mu_s) := \prod_{i=1}^j R_{si}(\mu_s), \; j \in \mathbb{N},
\end{equation}
is a test supermartingale for stratum $s$ for the composite null $\theta_s \le \mu_s$, $s=1, \ldots, S$.

We will now assemble the intersection test supermartingale by multiplying terms from
different test supermartingales for individual strata, in an order that can be chosen adaptively.
The \emph{stratum selector} $S(i) : \mathbb{N} \rightarrow \{1, \ldots, S\}$ is the stratum from which the $i$th 
term in the intersection test supermartingale will come.
The stratum selector $S(\cdot)$ can depend \emph{predictably} on the sample: it can 
depend on $(X_{S(j)J(j)})_{j=1}^{i-1}$ but not on $X_{S(k)J(k)}$ for $k \ge j$.
One example stratum selector is round-robin, $S(i) = (i \mod S)+1$, skipping any strata
that have been exhausted.
Another example concatenates the samples across strata: if we have drawn $n_s$ times from stratum $s$,
then $S(i) = 1$, $1 \le i \le n_1$; $S(i) = 2$, $n_1 < i \le n_1+n_2$; etc.

At time $i$, the intersection test supermartingale includes $J(i):= \#\{j \le i : S(j)=S(i)\}$ terms from stratum $S(i)$;
$S(j+1)$ and $J(j+1)$ are predictable from $\{X_{S(i)J(i)}\}_{i=1}^j$.
With this notation, the intersection test supermartingale is:
\begin{equation}
T_j(\boldsymbol{\mu}) := \prod_{i=1}^j R_{S(i)J(i)}(\mu_{S(i)}).
\end{equation}
Suppose $S(j+1) = s$ and $J(j+1) = \ell$.
Samples from different strata are independent, so 
the conditional expectation of $R_{S(j+1)J(j+1)}(\mu_{S(j+1)})$ given $\{X_{S(i)J(i)}\}_{i=1}^j$ is the conditional expectation of
$R_{s\ell}(\mu_s)$ given $\{X_{si}\}_{i=1}^{\ell-1}$, computed on the assumption that $\theta_s = \mu_s$, which is at most 1.
Thus 
\begin{equation}
\EE \left (T_{j+1}(\boldsymbol{\mu}) | (X_{S(i)J(i)})_{i=1}^j \right ) = \EE \left (T_j(\boldsymbol{\mu}) R_{S(j+1)J(j+1)}(\mu_{S(j+1)}) | (X_{S(i)J(i)})_{i=1}^j \right ) \le T_j(\boldsymbol{\mu}).
\end{equation}
That is, under the intersection null, $(T_j(\boldsymbol{\mu}))_{j \in \mathbb{N}}$ is a nonnegative supermartingale starting at 1,
and by Ville's inequality,
\begin{equation}
    \Pr (\max_j T_j(\boldsymbol{\mu}) \ge 1/\alpha) \le \alpha
\end{equation}
if $\boldsymbol{\theta} \le \boldsymbol{\mu}$.

In general, the power of the test of the intersection null will depend on the stratum selector $S(\cdot)$,
which can be adaptive.
For instance, if data from stratum $s$ suggest that $\theta_s \le \mu_s$, future values of $S(i)$ might omit stratum $s$
or sample from $s$ less frequently,
instead sampling preferentially from strata where there is some evidence that the intersection null is false, to maximize the
expected rate at which the test supermartingale grows, minimizing the $P$-value.
Indeed, for a fixed $\boldsymbol{\mu}$, choosing $S(i)$ can be viewed as a (possibly finite-population) multi-armed bandit problem:
which stratum should the next sample come from to maximize the expected rate of growth of the 
test statistic?
An additional complication is that we want fast growth for \emph{all} vectors $\boldsymbol{\mu}$ of stratumwise
means for which the population mean $\tilde{\boldsymbol{\mu}} := N^{-1}\sum_s N_s \mu_s \le 1/2$.
Importantly, different stratum selectors can be used for different values of $\boldsymbol{\mu}$;
this flexibility is explored by \citet{spertusStark22}.

To audit a given assertion, we need to check whether there is any $\boldsymbol{\mu} = (\mu_1, \ldots, \mu_S) \in [0, u]^S$ with 
$\tilde{\boldsymbol{\mu}}  \le 1/2$ for which
$\max_j T_j(\boldsymbol{\mu}) < 1/\alpha$.
If there is, sampling needs to continue.
We thus need to find
\begin{equation} \label{eq:strat-max}
  P_j^S := \max_{\boldsymbol{\mu} \in [0, u]^S: \tilde{\boldsymbol{\mu}} \le 1/2} (\max_j \; T_j(\boldsymbol{\mu}))^{-1},
\end{equation}
the solution to a finite-dimensional optimization problem.

\section{Bernoulli Sampling} \label{sec:bernoulli}

\citet{ottoboniEtal20} developed a ballot-polling risk-limiting audit (BBP) based on Bernoulli sampling, where each
ballot card is independently included in the sample with probability $p$.
This results in a random sample of random size.
Conditional on the attained sample size, it is a simple random sample of ballot cards.
Their approach to testing whether one candidate received more votes than another 
involves conditioning on the attained sample size and maximizing an SPRT $P$-value over
a nuisance parameter, the number of ballot cards that do not contain a valid vote for either of those candidates.
They find that BBP requires sample sizes comparable to BRAVO for the same margin.

ALPHA, combined with the SHANGRLA transformation, eliminates the need to perform the maximization
over a nuisance parameter.
To use ALPHA, the sample needs to have a notional ordering.
That ordering can come from randomly permuting the sample, or from setting a
canonical ordering of the ballot cards before the sample is selected, e.g., a lexicographical ordering,
then considering the sample order to be the lexicographical order of the cards in the sample.

If the initial Bernoulli sample does not suffice to confirm the outcome, the sample can be expanded
using the approach in \citet[Section 4]{ottoboniEtal20}.
Since the performance of BBP is similar to that of BRAVO, one might
expect that applying ALPHA to Bernoulli samples would require lower sample sizes (i.e., lower selection probabilities)
than BBP.
We do not perform any simulations here, but we plan to investigate the efficiency of ALPHA/SHANGRLA
versus BBP in future work.

\section{Simulations}

\subsection{Sampling with replacement}
Table~\ref{tab:results} reports mean sample sizes of ALPHA and BRAVO for the same true vote shares $\theta$,
with the same choices of $\eta$, using the truncated shrinkage estimate of $\eta_i$, for a variety of choices of $d$,
all for a risk limit of 5\%.
Results are based on 1,000 replications for each true $\theta$.
Sample sizes were limited to 10~million ballot cards: if a method required a sample bigger than that in any of the 1,000
replications, the result is listed as `---'.
As expected, BRAVO is best (or tied for best) when $\eta = \theta$, i.e., when the reported vote shares are exactly right.
ALPHA with a small value of $d$ is best when $|\eta - \theta|$ is large; and ALPHA with a small value of $d$ is
best when $|\eta - \theta|>0$ is small, except in a few cases where BRAVO beats ALPHA with $d=1000$ when $\theta > \eta$ and
$|\theta - \eta|$ is small.
When $\theta$ is large, ALPHA often does as well as BRAVO even when $\theta=\eta$.
When vote shares are wrong---and the reported winner still won---ALPHA often reduces
average sample sizes substantially, even when the true margin is large.
Indeed, in many cases, the sample size for BRAVO exceeded 10~million ballot cards in some runs, while the average for ALPHA was
up to five orders of magnitude lower.

The SPRT is known to perform poorly---sometimes never leading to a decision---when $\mu < \theta < \eta$.
In such cases, ALPHA did much better for all choices of $d$ when $\theta \ge 0.51$, and for $d= 10$ and $d=100$
when $\theta=0.505$.

The simulations show that the performance of BRAVO can also be poor when $\eta < \theta$.
In most of those cases, ALPHA performed better than BRAVO for all choices of $d$.
For instance, when $\theta=0.6$ (a margin of 20\%) and $\eta=0.7$,  ALPHA mean sample sizes were 204--353, but BRAVO sample sizes
exceeded $10^7$ for some runs.

\begin{table}
\centering
\tiny
\begin{tabular}{lr|rrrr|r}
  \multicolumn{2}{c|}{}    & \multicolumn{4}{|c|}{ALPHA with $d=$} & \multicolumn{1}{c}{} \\ 
\multicolumn{1}{c}{$\theta$}     &  \multicolumn{1}{c|}{$\eta$}   &  10 & 100 & 500 & 1000 & \multicolumn{1}{c}{BRAVO} \\
\hline
0.505 & 0.505 & 102,500 & 91,024 & 82,757 & 79,414 & \bf{58,266} \\
      & 0.51 & 102,738 & 91,878 & 84,088 & \bf{80,564} & --- \\
      & 0.52 & 103,418 & 93,842 & 88,512 & \bf{87,685} & --- \\
      & 0.53 & 103,746 & \bf{96,611} & 97,731 & 103,630 & --- \\
      & 0.54 & 104,490 & \bf{99,535} & 110,216 & 126,618 & --- \\
      & 0.55 & 105,071 & \bf{104,047} & 125,659 & 158,247 & --- \\
      & 0.6 & \bf{110,346} & 135,445 & 269,961 & 440,573 & --- \\
      & 0.65 & \bf{118,727} & 190,702 & 519,166 & 920,839 & --- \\
      & 0.7 & \bf{129,332} & 265,380 & 861,560 & 1,597,004 & --- \\
\hline
0.51 & 0.505 & 24,476 & 21,487 & 19,798 & \bf{19,258} & 19,965 \\
     & 0.51 & 24,598 & 21,598 & 19,577 & 18,841 & \bf{14,930} \\
     & 0.52 & 24,717 & 22,036 & 19,839 & \bf{19,035} & --- \\
     & 0.53 & 24,760 & 22,451 & \bf{20,846} & 20,928 & --- \\
     & 0.54 & 24,930 & 23,029 & \bf{22,888} & 24,602 & --- \\
     & 0.55 & 25,017 & \bf{23,856} & 25,848 & 30,351 & --- \\
     & 0.6 & \bf{25,954} & 30,041 & 57,261 & 93,849 & --- \\
     & 0.65 & \bf{27,721} & 42,078 & 116,207 & 209,576 & --- \\
     & 0.7 & \bf{30,117} & 61,550 & 201,768 & 376,304 & --- \\
\hline
0.52 & 0.505 & 5,531 & 4,944 & \bf{4,699} & 4,797 & 8,424 \\
     & 0.51 & 5,547 & 4,889 & 4,551 & \bf{4,490} & 4,959 \\
     & 0.52 & 5,584 & 4,882 & 4,291 & \bf{4,127} & 3,590 \\
     & 0.53 & 5,594 & 4,854 & 4,287 & \bf{4,091} & 4,583 \\
     & 0.54 & 5,631 & 4,898 & 4,400 & \bf{4,352} & --- \\
     & 0.55 & 5,660 & 4,996 & \bf{4,732} & 4,941 & --- \\
     & 0.6 & \bf{5,797} & 6,020 & 9,973 & 15,609 & --- \\
     & 0.65 & \bf{6,165} & 8,460 & 22,847 & 41,464 & --- \\
     & 0.7 & \bf{6,628} & 12,620 & 42,554 & 79,707 & --- \\
\hline
0.53 & 0.505 & 2,447 & \bf{2,238} & 2,321 & 2,441 & 5,433 \\
     & 0.51 & 2,452 & 2,220 & \bf{2,216} & 2,259 & 3,013 \\
     & 0.52 & 2,464 & 2,204 & 2,058 & 1,991 & \bf{1,911} \\
     & 0.53 & 2,473 & 2,192 & 1,964 & 1,852 & \bf{1,717} \\
     & 0.54 & 2,489 & 2,189 & 1,942 & \bf{1,852} & 1,946 \\
     & 0.55 & 2,512 & 2,196 & 1,993 & \bf{1,987} & 3,076 \\
     & 0.6 & 2,600 & \bf{2,551} & 3,556 & 4,991 & --- \\
     & 0.65 & \bf{2,748} & 3,468 & 8,206 & 14,617 & --- \\
     & 0.7 & \bf{2,988} & 5,079 & 16,505 & 30,741 & --- \\
\hline
0.54 & 0.505 & 1,326 & \bf{1,244} & 1,384 & 1,525 & 4,023 \\
     & 0.51 & 1,329 & \bf{1,233} & 1,293 & 1,384 & 2,141 \\
     & 0.52 & 1,323 & 1,201 & \bf{1,162} & 1,169 & 1,257 \\
     & 0.53 & 1,322 & 1,181 & 1,092 & 1,060 & \bf{1,011} \\
     & 0.54 & 1,329 & 1,167 & 1,046 & 1,001 & \bf{953} \\
     & 0.55 & 1,337 & 1,162 & 1,034 & \bf{1,009} & 1,048 \\
     & 0.6 & 1,369 & \bf{1,247} & 1,524 & 1,910 & --- \\
     & 0.65 & \bf{1,443} & 1,670 & 3,507 & 5,791 & --- \\
     & 0.7 & \bf{1,544} & 2,422 & 7,464 & 14,263 & --- \\
\hline
0.55 & 0.505 & 820 & \bf{801} & 959 & 1,102 & 3,172 \\
     & 0.51 & 820 & \bf{788} & 891 & 982 & 1,688 \\
     & 0.52 & 816 & \bf{763} & 773 & 809 & 943 \\
     & 0.53 & 819 & 737 & 700 & \bf{695} & 707 \\
     & 0.54 & 820 & 711 & 650 & 635 & \bf{618} \\
     & 0.55 & 822 & 702 & 626 & 608 & \bf{598} \\
     & 0.6 & 826 & \bf{737} & 799 & 898 & --- \\
     & 0.65 & \bf{873} & 945 & 1,701 & 2,655 & --- \\
     & 0.7 & \bf{950} & 1,402 & 4,028 & 7,434 & --- \\
\hline
0.6 & 0.505 & \bf{195} & 235 & 345 & 426 & 1,529 \\
    & 0.51 & \bf{195} & 227 & 316 & 378 & 783 \\
    & 0.52 & \bf{193} & 214 & 271 & 304 & 413 \\
    & 0.53 & \bf{191} & 202 & 236 & 253 & 292 \\
    & 0.54 & \bf{191} & 192 & 209 & 218 & 234 \\
    & 0.55 & \bf{191} & 183 & 191 & 194 & 199 \\
    & 0.6 & 187 & 160 & 151 & 151 & \bf{149} \\
    & 0.65 & 189 & \bf{163} & 170 & 175 & 198 \\
    & 0.7 & \bf{195} & 197 & 271 & 326 & --- \\
\hline
0.65 & 0.505 & \bf{9}1 & 127 & 207 & 264 & 1,016 \\
     & 0.51 & \bf{91} & 123 & 190 & 234 & 515 \\
     & 0.52 & \bf{90} & 115 & 161 & 187 & 267 \\
     & 0.53 & \bf{89} & 108 & 139 & 153 & 185 \\
     & 0.54 & \bf{88} & 102 & 122 & 130 & 144 \\
     & 0.55 & \bf{87} & 96 & 110 & 114 & 121 \\
     & 0.6 & 84 & \bf{77} & \bf{77} & \bf{77} & \bf{77} \\
     & 0.65 & 82 & 71 & \bf{69} & \bf{69} & \bf{69} \\
     & 0.7 & 83 & \bf{73} & 75 & 76 & 79 \\
\hline
0.7 & 0.505 & \bf{54} & 85 & 146 & 189 & 757 \\
     & 0.51 & \bf{54} & 82 & 134 & 167 & 384 \\
     & 0.52 & \bf{53} & 76 & 113 & 133 & 196 \\
     & 0.53 & \bf{52} & 71 & 98 & 109 & 135 \\
     & 0.54 & \bf{51} & 66 & 85 & 92 & 104 \\
     & 0.55 & \bf{51} & 62 & 75 & 80 & 86 \\
     & 0.6 & \bf{47} & 49 & 51 & 51 & 51 \\
     & 0.65 & 45 & 42 & 42 & 42 & 41 \\
     & 0.7 & 45 & 40 & \bf{39} & \bf{39} & \bf{40} 
\end{tabular}
\caption{\protect \label{tab:results} Estimated sample sizes for a ballot-polling audit using sampling with replacement
to confirm the outcome at a risk limit of $0.05$, for ALPHA (with a variety of choices of
$d$) versus BRAVO.
$\theta$: actual vote share for winner.
$\eta$: reported vote share for winner.
Average of 1,000 replications.
``--'' indicates that in at least one replication, the sample size exceeded 10~million.
For each $\theta$, $\eta$ pair, the smallest average sample size is in bold font.
}
\end{table}

\subsection{Sampling without replacement}
Table~\ref{tab:finite-population} compares several methods for ballot-polling without replacement, again in a two-candidate
plurality contest with no invalid votes or votes for other candidates, based on $10^5$ pseudo-random audits
for each hypothetical population.
The methods listed include the best-performing method in RiLACS \citep{waudby-smithEtal21} 
that uses an explicit alternative $\eta$ (\emph{a priori} Kelly), the best-performing method in RiLACS
that does not use a pre-specified alternative (SqKelly), 
Wald's SPRT for sampling without replacement (the analogue of BRAVO for sampling without replacement), and
ALPHA with the truncated shrinkage estimator for a variety of values of
$d$. 
Methods that use an explicit alternative (\emph{a priori} Kelly, SPRT, ALPHA) were tested using a range of values of $\eta$.
Kaplan's martingale \citep{stark20} was not included because it is expensive to compute, numerically unstable, and
performs comparably to some of the methods studied by \citet{waudby-smithEtal21}, such as dKelly.
The election parameters $N$, $\theta$, and $\eta$ were chosen to make the simulations commensurable with 
\citet{huangEtal20}.

The columns labeled $n \le 2{,}000$ are for an audit that examines up to 2,000 ballots selected at random, and
if the outcome has not been confirmed by then, examines the remaining 18,000 ballots to determine who won.
This is consistent with how RLAs may be conducted in practice: retrieving randomly selected ballots has a fair amount of 
overhead, so there is a sampling fraction above which it is more efficient to examine every ballot rather than to sample cards
at random.
That threshold depends on how ballots are organized, the size of storage batches, whether the ballot cards have been
imprinted with identifiers, etc., but based on experience, the break-even point 
is when the sample size reaches 5--15\% of the
population size.
Thus, for a population of $N=20{,}000$ ballot cards, an election official might elect to conduct a full hand count
if the audit sample becomes larger than 2,000 cards, 10\% of the population. 
The entries for Bayesian, BRAVO, and ClipAudit are derived from Table~2 of \citet{huangEtal20} by adding 
$18,000 \times (1-\mbox{power})$ to the entries, since that table was calculated by capping the sample size
at $2,000$.

The columns in Table~\ref{tab:finite-population}  for $n \le N = 20{,}000$ do not restrict the sample size;
they show the expected sample size of audits when the audit is allowed to escalate one card at a time,
potentially to a full hand count.
Because the sample is drawn without replacement from a population of size 20,000, the audits are guaranteed to reject the
null hypothesis by the time the sample size is 20,000.

Many of the methods perform comparably. 
SqKelly is often best when $\theta$ is not equal to $\eta$ for any of the methods that
use $\eta$.
\emph{A priori} Kelly with $\eta=0.7$, the SPRT with $\eta=0.7$, and ALPHA with $\eta=0.7$ and $d=1{,}000$
work relatively well against a broad range of alternatives.
ALPHA is broadly competitive, despite the fact that no effort has gone into optimizing the estimator $\eta_i$ to minimize
the expected sample size.

\begin{table}
\centering
\tiny
\begin{tabular}{ll|rrrrrrr|rrrrrrr}& & \multicolumn{7}{|c|}{$n \le2{,}000$} &  \multicolumn{7}{|c}{$n \le N=20{,}000$} \\ 
Method & & \multicolumn{7}{c|}{mean sample size, $\theta=$}  & \multicolumn{7}{|c}{mean sample size, $\theta=$ }\\ 
& & .505 & .51 & .52 & .55 & .6 & .64 & .7  & .505 & .51 & .52 & .55 & .6 & .64 & .7  \\ 
\hline
sqKelly  &  & 18,401  & 17,224  & 12,881  & \bf{813}  & \bf{181}  & 110  & 68   & 17,917  & 14,255  & 4,844  & 587  & \bf{181}  & 110  & 68  \\ 
\hline
a priori Kelly & $\eta=0.51$   & 19,985  & 19,931  & 19,288  & 4,234  & 774  & 548  & 381   & 13,823  & 8,351  & 4,195  & 1,591  & 774  & 548  & 381  \\ 
 & $\eta=0.55$   & \bf{18,350}  & \bf{17,161}  & \bf{12,848}  & 823  & 200  & 131  & 86   & 18,049  & 14,922  & 5,447  & 578  & 200  & 131  & 86  \\ 
 & $\eta=0.7$   & 19,004  & 18,839  & 18,449  & 16,064  & 2,821  & 98  & \bf{38}   & 18,818  & 18,472  & 17,742  & 12,937  & 703  & 98  & \bf{38}  \\ 
\hline
 SPRT & $\eta=0.51$  & 19,936  & 19,758  & 18,243  & 2,620  & 671  & 475  & 329   & \bf{13,085}  & \bf{7,702}  & 3,751  & 1,392  & 671  & 475  & 329  \\  
 & $\eta=0.55$  & \bf{18,350}  & 17,181  & 12,910  & 832  & 199  & 130  & 85   & 18,028  & 15,762  & 6,333  & \bf{578}  & 199  & 130  & 85  \\  
 & $\eta=0.7$  & 19,005  & 18,840  & 18,451  & 16,076  & 3,189  & 99  & \bf{38}   & 18,818  & 18,472  & 17,743  & 14,260  & 881  & 98  & \bf{38}  \\  
\hline
ALPHA & $\eta=0.51$ $d=10$ & 19,130  & 18,475  & 15,504  & 1,373  & 197  & 102  & 52  & 14,841  & 9,464  & 4,032  & 780  & 197  & 102  & 52  \\ 
 & $\eta=0.51$ $d=100$ & 19,220  & 18,431  & 14,807  & 1,121  & 227  & 135  & 81  & 14,406  & 8,888  & 3,677  & 751  & 227  & 135  & 81  \\ 
 & $\eta=0.51$ $d=500$ & 19,397  & 18,603  & 14,821  & 1,152  & 313  & 204  & 133  & 14,096  & 8,508  & 3,533  & 840  & 313  & 204  & 133  \\ 
 & $\eta=0.51$ $d=1{,}000$ & 19,505  & 18,786  & 15,140  & 1,250  & 371  & 248  & 165  & 13,936  & 8,343  & \bf{3,512}  & 918  & 371  & 248  & 165  \\ 
 & $\eta=0.55$ $d=10$ & 19,078  & 18,440  & 15,568  & 1,407  & 192  & 98  & 49  & 14,937  & 9,578  & 4,089  & 780  & 192  & 98  & 49  \\ 
 & $\eta=0.55$ $d=100$ & 18,892  & 18,034  & 14,429  & 1,052  & 184  & 105  & 62  & 14,716  & 9,195  & 3,726  & 676  & 184  & 105  & 62  \\ 
 & $\eta=0.55$ $d=500$ & 18,576  & 17,492  & 13,274  & 857  & 190  & 118  & 75  & 15,032  & 9,357  & 3,538  & 609  & 190  & 118  & 75  \\ 
 & $\eta=0.55$ $d=1{,}000$ & 18,473  & 17,311  & 12,989  & 823  & 193  & 123  & 79  & 15,571  & 9,880  & 3,622  & 594  & 193  & 123  & 79  \\ 
 & $\eta=0.7$ $d=10$ & 19,041  & 18,602  & 16,547  & 1,926  & 196  & 93  & 43  & 15,696  & 10,563  & 4,685  & 874  & 196  & 93  & 43  \\ 
 & $\eta=0.7$ $d=100$ & 18,991  & 18,753  & 17,929  & 4,957  & 199  & \bf{85}  & \bf{38}  & 17,497  & 13,807  & 7,189  & 1,221  & 199  & \bf{85}  & \bf{38}  \\ 
 & $\eta=0.7$ $d=500$ & 18,985  & 18,815  & 18,387  & 14,085  & 275  & 89  & \bf{38}  & 18,537  & 17,088  & 12,656  & 2,961  & 271  & 89  & \bf{38}  \\ 
 & $\eta=0.7$ $d=1{,}000$ & 18,993  & 18,824  & 18,416  & 15,544  & 392  & 92  & \bf{38}  & 18,731  & 17,844  & 14,811  & 4,692  & 327  & 92  & \bf{38}  \\ 
\hline
Bayesian & $a=b=1$ & & 18,669 & 16,794  & 2,148  & 198  & 95  & 44  \\
\hline
BRAVO & $\eta=0.51$ &&19,962 & 19,525  & 5,133  & 790  & 556  & 385  \\
             & $\eta=0.55$ && 17,408 & 13,371  & 932  & 200  & 131  & 86  \\
             & $\eta=0.7$   && 18,844& 18,433  & 16,021  & 3,612  & 99  & \bf{38}  \\
\hline
ClipAudit &      && 17,462 & 13,547  & 913  & 167  & 88  & 45  \\
\hline
\end{tabular} 
\caption{\protect \label{tab:finite-population}
Estimated workload for ballot-polling audits using sampling without replacement from a population of 
size 20,000 at risk limit 5\%.
Because the sample is drawn without replacement, all these methods are guaranteed to reject the null hypothesis by the time
the sample size is 20,000, if the null is false.
`SqKelly' does not require an explicit alternative value for $\theta$; it optimizes against a mixture of possibilities that assigns higher weight
to smaller margins.
`\emph{A priori} Kelly' is the betting martingale that maximizes the expected rate of growth of the test statistic when $\theta = \eta$.
Samples are drawn without replacement from a population of size 20,000 of which a fraction $\theta$ are 1 and a fraction $(1-\theta)$
are zero, so the population mean is $\theta$.
SPRT is Wald's sequential probability ratio test for sampling without replacement from a binary population, the ``without-replacement''
version of the test BRAVO uses.
It is equivalent to ALPHA using the estimate $\hat{\theta}_j = \eta$.
Under the null hypothesis, $\theta=1/2$. 
ALPHA, \emph{a priori} Kelly, and the SPRT use an alternative value, $\eta > 1/2$, such as the reported population mean. 
Results reflect $10^5$ simulations for each value of $\theta$.
Columns 3:9 are mean sample sizes for an audit that samples at most 2,000 ballot cards before proceeding to
a full hand count of all 20,000 ballot cards if the outcome has not been confirmed by then. 
Columns 10:16 give the mean sample sizes to reject the null when the sample is allowed to
expand to comprise the whole population of 20,000 ballot cards.
The five bottom rows are from supplementary materials in \citet{huangEtal20} available at 
\url{https://github.com/dvukcevic/AuditAnalysis/blob/master/combined_tables/n\%3D020000_m\%3D02000_p\%3D0.500_replacement\%3DFalse_step\%3D1/unconditional_mean_with_recount_addin.csv\#L27} (last visited 1~February 2022).
The ``Bayesian'' test uses a risk-maximizing prior \citep{vora19}, in this case, a point mass at $1/2$ mixed with a uniform on $(1/2, 1]$, 
which makes it risk-limiting.
ClipAudit \citep{rivest17} is calibrated in a way that \emph{almost} limits the risk; in simulations it was $5.1\%$ \citep{huangEtal20}.
The smallest average sample size in each column, omitting ClipAudit, is in bold font.
}
\end{table}

\subsection{Sampling without replacement when some ballot cards do not have a valid vote for either candidate} \label{sec:blanks}

As discussed above, BRAVO relies on testing conditional probabilities
rather than unconditional probabilities when there are ballot cards with no valid vote for either candidate.
For sampling without replacement (and for stratified sampling), that approach does not work.
Here, we estimate expected sample sizes for sampling without replacement from populations of different sizes
with different fractions of ballot cards with no valid vote for either candidate, using the SHANGRLA assorter for plurality
contests described in section~\ref{sec:shangrla}, which assigns such ballot cards the value $1/2$.
Tables~\ref{tab:without-replacement-blanks-1} and \ref{tab:without-replacement-blanks-2} show the
 results for the tests that can work with nonbinary data:
the SqKelly and \emph{a priori} Kelly martingales, and the ALPHA supermartingales. 
Kaplan's martingale, the Kaplan-Wald test, and the Kaplan-Kolmogorov test \citep{stark09b,stark20} could also be used, but we do not explore their performance here.
(However, see section \ref{sec:comparison-audits}.)

The smallest expected sample sizes are generally for \emph{a priori} Kelly with $\eta=\theta$, with non-adaptive ALPHA (corresponding to
$d=\infty$) nearly tied and sometimes winning.
SqKelly does nearly as well when $\theta \ge 0.55$.

\begin{table}
\centering
\tiny
\begin{tabular}{lll|rrrr|rrrr|rrrr} 
& & & \multicolumn{4}{|c|}{$N=10{,}000$, \%blank} &  \multicolumn{4}{|c|}{$N=100{,}000$, \%blank} & \multicolumn{4}{|c}{$N=500{,}000$, \%blank} \\ 
$\theta$ & Method & params & 10 & 25 & 50 & 75  & 10 & 25 & 50 & 75  & 10 & 25 & 50 & 75  \\
\hline 0.51 & sqKelly & & 7,232  & 7,179  & 7,703  & 8,131  & 68,433  & 69,862  & 70,274  & 69,601  & 354,731  & 356,437  & 355,429  & 354,966  \\
\cline{2-15} & apKelly & $\eta=$0.51 & 6,452  & 6,919  & 8,138  & 9,519  & \bf{13,493}  & \bf{15,490}  & 21,474  & 35,699  & \bf{15,914}  & \bf{18,982}  & \bf{27,024}  & 50,081  \\
\cline{2-15}
& ALPHA & $\eta=$0.51 $d=$10 & 6,265  & 6,523  & 7,655  & 9,125  & 19,031  & 21,443  & 30,361  & 56,160  & 24,289  & 28,367  & 45,379  & 112,638  \\
&  & $\eta=$0.51 $d=$100 & 6,034  & 6,449  & 7,669  & 9,150  & 17,228  & 20,395  & 29,878  & 56,006  & 22,079  & 26,755  & 44,259  & 112,125  \\
&  & $\eta=$0.51 $d=$1000 & 5,889  & 6,387  & 7,690  & 9,198  & 15,940  & 19,301  & 29,607  & 56,319  & 20,164  & 24,879  & 43,144  & 112,242  \\
 &  & $\eta=$0.51 $d=\infty$ & 5,736  & 6,294  & 7,754  & 9,243  & 13,514  & 16,645  & 27,832  & 56,680  & 16,129  & 20,382  & 36,289  & 107,164  \\
\cline{2-15} & apKelly & $\eta=$0.52 & \bf{5,541}  & \bf{5,841}  & 6,962  & 8,469  & 17,077  & 19,236  & 24,147  & 34,630  & 33,810  & 40,384  & 49,785  & 73,143  \\
\cline{2-15}
& ALPHA & $\eta=$0.52 $d=$10 & 6,266  & 6,527  & 7,654  & 9,123  & 19,077  & 21,447  & 30,354  & 56,125  & 24,338  & 28,395  & 45,362  & 112,526  \\
&  & $\eta=$0.52 $d=$100 & 6,025  & 6,410  & 7,653  & 9,142  & 17,385  & 20,327  & 29,714  & 55,896  & 22,265  & 26,824  & 44,098  & 111,701  \\
&  & $\eta=$0.52 $d=$1000 & 5,750  & 6,182  & 7,554  & 9,164  & 15,412  & 18,621  & 28,611  & 55,566  & 19,659  & 24,007  & 41,625  & 110,018  \\
 &  & $\eta=$0.52 $d=\infty$ & 5,367  & 5,710  & 7,178  & 9,089  & 15,687  & 15,682  & \bf{21,054}  & 45,197  & 26,109  & 23,260  & 27,070  & 68,370  \\
\cline{2-15} & apKelly & $\eta=$0.55 & 7,447  & 7,401  & 7,847  & 8,177  & 72,111  & 73,492  & 74,529  & 73,563  & 370,856  & 373,252  & 372,443  & 379,728  \\
\cline{2-15}
& ALPHA & $\eta=$0.55 $d=$10 & 6,277  & 6,544  & 7,653  & 9,113  & 19,369  & 21,536  & 30,278  & 56,013  & 24,582  & 28,603  & 45,336  & 112,173  \\
&  & $\eta=$0.55 $d=$100 & 6,096  & 6,393  & 7,610  & 9,122  & 18,079  & 20,446  & 29,416  & 55,617  & 23,388  & 27,360  & 43,923  & 110,973  \\
&  & $\eta=$0.55 $d=$1000 & 6,062  & 6,081  & 7,263  & 9,042  & 18,948  & 19,844  & 26,694  & 53,281  & 26,062  & 27,117  & 39,394  & 104,207  \\
 &  & $\eta=$0.55 $d=\infty$ & 7,430  & 6,714  & \bf{6,728}  & 8,535   & 70,863  & 62,934  & 32,088  & 33,206  & 368,165  & 320,200  & 113,614  & \bf{49,577}  \\
\cline{2-15} & apKelly & $\eta=$0.6 & 8,899  & 8,951  & 9,080  & 9,053  & 89,799  & 88,007  & 88,745  & 90,466  & 440,987  & 441,420  & 451,707  & 455,784  \\
\cline{2-15}
& ALPHA & $\eta=$0.6 $d=$10 & 6,350  & 6,558  & 7,653  & 9,101  & 19,836  & 21,672  & 30,299  & 55,872  & 25,447  & 29,131  & 45,331  & 111,736  \\
&  & $\eta=$0.6 $d=$100 & 6,397  & 6,489  & 7,538  & 9,081  & 20,735  & 21,801  & 29,441  & 55,112  & 27,540  & 30,299  & 44,542  & 109,761  \\
&  & $\eta=$0.6 $d=$1000 & 7,672  & 7,112  & 7,153  & 8,840  & 38,270  & 30,979  & 28,531  & 50,166  & 62,351  & 49,420  & 44,781  & 97,482  \\
 &  & $\eta=$0.6 $d=\infty$ & 8,829  & 8,675  & 7,881  & \bf{7,901}  & 88,324  & 86,296  & 75,852  & \bf{39,441}  & 436,073  & 418,900  & 379,683  & 124,103  \\
\cline{2-15} & apKelly & $\eta=$0.7 & 9,208  & 9,296  & 9,372  & 9,267  & 92,733  & 92,418  & 92,482  & 92,947  & 449,032  & 448,176  & 458,981  & 462,335  \\
\cline{2-15}
& ALPHA & $\eta=$0.7 $d=$10 & 6,608  & 6,710  & 7,620  & 9,081  & 21,443  & 22,691  & 30,271  & 55,517  & 27,815  & 30,865  & 45,644  & 110,818  \\
&  & $\eta=$0.7 $d=$100 & 7,506  & 7,171  & 7,525  & 8,979  & 32,071  & 28,929  & 31,015  & 54,188  & 46,341  & 42,398  & 48,478  & 108,507  \\
&  & $\eta=$0.7 $d=$1000 & 8,951  & 8,791  & 8,043  & 8,463  & 69,941  & 60,976  & 43,594  & 48,143  & 177,722  & 136,461  & 82,340  & 96,717  \\
 &  & $\eta=$0.7 $d=\infty$ & 9,101  & 9,210  & 9,049  & 8,111  & 92,073  & 90,313  & 88,248  & 74,155  & 456,027  & 448,199  & 439,668  & 383,048  \\
\hline 0.52 & sqKelly & & 3,417  & 3,617  & 4,262  & 5,483  & 13,187  & 14,715  & 17,276  & 21,988  & 45,054  & 43,688  & 51,205  & 58,535  \\
\cline{2-15} & apKelly & $\eta=$0.51 & 3,935  & 4,450  & 5,775  & 7,993  & 5,328  & 6,492  & 9,445  & 17,432  & 5,530  & 6,443  & 9,755  & 19,140  \\
\cline{2-15}
& ALPHA & $\eta=$0.51 $d=$10 & 3,420  & 3,833  & 5,034  & 7,552  & 5,643  & 6,776  & 11,211  & 26,830  & 6,178  & 7,038  & 12,293  & 35,333  \\
&  & $\eta=$0.51 $d=$100 & 3,280  & 3,730  & 5,043  & 7,574  & 5,244  & 6,479  & 11,048  & 26,859  & 5,595  & 6,718  & 12,028  & 35,271  \\
&  & $\eta=$0.51 $d=$1000 & 3,211  & 3,737  & 5,189  & 7,693  & 5,057  & 6,407  & 11,301  & 27,495  & 5,378  & 6,592  & 12,306  & 36,092  \\
 &  & $\eta=$0.51 $d=\infty$ & 3,402  & 4,057  & 5,673  & 8,006  & 5,527  & 7,479  & 14,148  & 35,146  & 5,912  & 7,821  & 16,326  & 53,585  \\
\cline{2-15} & apKelly & $\eta=$0.52 & 2,866  & \bf{3,198}  & 4,202  & 6,172  & \bf{3,957}  & \bf{4,746}  & \bf{6,817}  & 12,439  & \bf{4,156}  & \bf{4,725}  & \bf{7,161}  & \bf{14,178}  \\
\cline{2-15}
& ALPHA & $\eta=$0.52 $d=$10 & 3,429  & 3,833  & 5,032  & 7,550  & 5,640  & 6,776  & 11,208  & 26,796  & 6,194  & 7,052  & 12,286  & 35,268  \\
&  & $\eta=$0.52 $d=$100 & 3,242  & 3,707  & 5,003  & 7,560  & 5,240  & 6,418  & 10,942  & 26,738  & 5,613  & 6,610  & 11,942  & 35,123  \\
&  & $\eta=$0.52 $d=$1000 & 3,017  & 3,475  & 4,972  & 7,604  & 4,656  & 5,884  & 10,531  & 26,734  & 4,864  & 5,989  & 11,475  & 34,816  \\
 &  & $\eta=$0.52 $d=\infty$ & \bf{2,819}  & 3,291  & 4,866  & 7,649  & 4,000  & 5,071  & 9,192  & 25,467  & 4,228  & 5,138  & 9,695  & 31,879  \\
\cline{2-15} & apKelly & $\eta=$0.55 & 3,612  & 3,817  & 4,274  & 5,459  & 18,933  & 19,814  & 22,049  & 25,739  & 84,685  & 80,473  & 86,120  & 87,629  \\
\cline{2-15}
& ALPHA & $\eta=$0.55 $d=$10 & 3,446  & 3,837  & 5,019  & 7,528  & 5,663  & 6,765  & 11,178  & 26,752  & 6,231  & 7,049  & 12,265  & 35,113  \\
&  & $\eta=$0.55 $d=$100 & 3,190  & 3,602  & 4,921  & 7,512  & 5,232  & 6,369  & 10,674  & 26,442  & 5,678  & 6,524  & 11,748  & 34,652  \\
&  & $\eta=$0.55 $d=$1000 & 2,967  & 3,205  & 4,448  & 7,346  & 4,878  & 5,583  & 8,982  & 24,501  & 5,228  & 5,534  & 9,920  & 31,626  \\
 &  & $\eta=$0.55 $d=\infty$ & 3,466  & 3,244  & \bf{3,799}  & 6,587  & 14,670  & 8,449  & 6,963  & 14,967  & 52,370  & 11,817  & 7,493  & 16,743  \\
\cline{2-15} & apKelly & $\eta=$0.6 & 7,508  & 7,508  & 7,429  & 7,592  & 75,380  & 75,596  & 76,025  & 75,706  & 370,396  & 381,595  & 402,570  & 385,000  \\
\cline{2-15}
& ALPHA & $\eta=$0.6 $d=$10 & 3,513  & 3,822  & 4,999  & 7,509  & 5,827  & 6,803  & 11,139  & 26,607  & 6,365  & 7,090  & 12,239  & 34,918  \\
&  & $\eta=$0.6 $d=$100 & 3,406  & 3,644  & 4,789  & 7,434  & 5,829  & 6,636  & 10,407  & 25,899  & 6,382  & 6,707  & 11,665  & 33,933  \\
&  & $\eta=$0.6 $d=$1000 & 4,536  & 3,971  & 4,174  & 6,895  & 10,609  & 8,389  & 8,708  & 21,683  & 12,148  & 8,825  & 9,965  & 27,422  \\
 &  & $\eta=$0.6 $d=\infty$ & 7,438  & 6,486  & 4,370  & 5,563  & 74,378  & 63,828  & 24,198  & \bf{12,325}  & 342,539  & 314,364  & 101,906  & 14,813  \\
\cline{2-15} & apKelly & $\eta=$0.7 & 8,829  & 8,784  & 8,921  & 9,114  & 87,007  & 88,622  & 90,646  & 93,576  & 411,057  & 414,569  & 435,188  & 452,283  \\
\cline{2-15}
& ALPHA & $\eta=$0.7 $d=$10 & 3,691  & 3,873  & 4,960  & 7,467  & 6,505  & 7,138  & 11,101  & 26,367  & 6,981  & 7,434  & 12,297  & 34,538  \\
&  & $\eta=$0.7 $d=$100 & 4,666  & 4,274  & 4,715  & 7,264  & 9,802  & 8,805  & 10,807  & 25,092  & 10,935  & 9,349  & 12,390  & 32,790  \\
&  & $\eta=$0.7 $d=$1000 & 7,746  & 6,865  & 5,179  & 6,226  & 35,622  & 25,820  & 14,518  & 19,018  & 51,880  & 33,540  & 17,704  & 23,960  \\
 &  & $\eta=$0.7 $d=\infty$ & 8,801  & 8,622  & 7,603  & \bf{5,455}  & 85,959  & 85,366  & 77,494  & 26,607  & 414,810  & 393,494  & 368,757  & 93,776  \\
\hline 0.55 & sqKelly & & 609  & 764  & 1,049  & 1,870  & 690  & 813  & 1,210  & 2,329  & 670  & 792  & \bf{1,198}  & 2,428  \\
\cline{2-15} & apKelly & $\eta=$0.51 & 1,669  & 1,957  & 2,764  & 4,720  & 1,842  & 2,198  & 3,285  & 6,423  & 1,844  & 2,200  & 3,343  & 6,613  \\
\cline{2-15}
& ALPHA & $\eta=$0.51 $d=$10 & 808  & 1,007  & 1,608  & 3,630  & 942  & 1,153  & 1,976  & 5,758  & 924  & 1,136  & 2,028  & 6,176  \\
&  & $\eta=$0.51 $d=$100 & 800  & 1,025  & 1,660  & 3,714  & 926  & 1,160  & 2,035  & 5,946  & 912  & 1,150  & 2,106  & 6,382  \\
&  & $\eta=$0.51 $d=$1000 & 984  & 1,243  & 1,997  & 4,123  & 1,137  & 1,446  & 2,529  & 6,920  & 1,136  & 1,439  & 2,626  & 7,409  \\
 &  & $\eta=$0.51 $d=\infty$ & 1,405  & 1,787  & 2,862  & 5,273  & 1,943  & 2,688  & 5,428  & 15,692  & 2,007  & 2,825  & 6,121  & 21,658  \\
\cline{2-15} & apKelly & $\eta=$0.52 & 972  & 1,163  & 1,663  & 2,970  & 1,051  & 1,244  & 1,838  & 3,655  & 1,046  & 1,240  & 1,893  & 3,704  \\
\cline{2-15}
& ALPHA & $\eta=$0.52 $d=$10 & 805  & 1,006  & 1,600  & 3,623  & 941  & 1,151  & 1,968  & 5,748  & 921  & 1,134  & 2,021  & 6,166  \\
&  & $\eta=$0.52 $d=$100 & 770  & 1,001  & 1,631  & 3,690  & 893  & 1,129  & 1,993  & 5,886  & 874  & 1,109  & 2,067  & 6,310  \\
&  & $\eta=$0.52 $d=$1000 & 844  & 1,100  & 1,822  & 3,983  & 950  & 1,239  & 2,272  & 6,555  & 953  & 1,237  & 2,357  & 7,020  \\
 &  & $\eta=$0.52 $d=\infty$ & 968  & 1,277  & 2,236  & 4,747  & 1,130  & 1,541  & 3,180  & 10,478  & 1,130  & 1,546  & 3,323  & 12,074  \\
\cline{2-15} & apKelly & $\eta=$0.55 & \bf{604}  & \bf{744}  & 1,036  & 1,856  & \bf{678}  & \bf{809}  & \bf{1,188}  & 2,297  & \bf{656}  & \bf{775}  & 1,207  & 2,377  \\
\cline{2-15}
& ALPHA & $\eta=$0.55 $d=$10 & 802  & 997  & 1,588  & 3,607  & 941  & 1,147  & 1,951  & 5,721  & 919  & 1,126  & 2,003  & 6,133  \\
&  & $\eta=$0.55 $d=$100 & 708  & 930  & 1,539  & 3,611  & 822  & 1,046  & 1,878  & 5,713  & 805  & 1,019  & 1,937  & 6,107  \\
&  & $\eta=$0.55 $d=$1000 & 637  & 841  & 1,440  & 3,586  & 716  & 907  & 1,712  & 5,603  & 691  & 887  & 1,767  & 5,930  \\
 &  & $\eta=$0.55 $d=\infty$ & 614  & 797  & 1,363  & 3,511  & 685  & 851  & 1,571  & 5,198  & 666  & 833  & 1,617  & 5,409  \\
\cline{2-15} & apKelly & $\eta=$0.6 & 1,089  & 1,243  & 1,531  & 2,270  & 2,901  & 3,159  & 3,805  & 5,552  & 5,349  & 6,175  & 7,098  & 11,932  \\
\cline{2-15}
& ALPHA & $\eta=$0.6 $d=$10 & 807  & 993  & 1,576  & 3,582  & 949  & 1,134  & 1,926  & 5,658  & 923  & 1,118  & 1,972  & 6,052  \\
&  & $\eta=$0.6 $d=$100 & 697  & 881  & 1,423  & 3,487  & 812  & 975  & 1,716  & 5,452  & 802  & 951  & 1,786  & 5,827  \\
&  & $\eta=$0.6 $d=$1000 & 704  & 787  & 1,144  & 3,031  & 838  & 870  & 1,338  & 4,401  & 846  & 852  & 1,340  & 4,584  \\
 &  & $\eta=$0.6 $d=\infty$ & 887  & 846  & \bf{1,030}  & 2,452  & 1,710  & 1,052  & \bf{1,188}  & 3,124  & 1,861  & 1,042  & 1,208  & 3,155  \\
\cline{2-15} & apKelly & $\eta=$0.7 & 6,530  & 6,496  & 6,541  & 6,528  & 65,591  & 66,245  & 67,839  & 66,608  & 305,925  & 315,550  & 338,236  & 364,794  \\
\cline{2-15}
& ALPHA & $\eta=$0.7 $d=$10 & 850  & 1,007  & 1,547  & 3,524  & 1,008  & 1,144  & 1,900  & 5,558  & 972  & 1,134  & 1,930  & 5,934  \\
&  & $\eta=$0.7 $d=$100 & 983  & 981  & 1,314  & 3,240  & 1,181  & 1,122  & 1,607  & 5,005  & 1,183  & 1,124  & 1,637  & 5,320  \\
&  & $\eta=$0.7 $d=$1000 & 2,642  & 1,661  & 1,203  & 2,338  & 4,231  & 2,321  & 1,487  & 3,147  & 4,594  & 2,414  & 1,493  & 3,287  \\
 &  & $\eta=$0.7 $d=\infty$ & 6,605  & 4,659  & 1,591  & \bf{1,840}  & 62,167  & 46,554  & 4,046  & \bf{2,295}  & 278,732  & 229,010  & 7,574  & \bf{2,376} 
\end{tabular} 
\caption{\protect \label{tab:without-replacement-blanks-1}
Estimated sample sizes for a ballot-polling audit using sampling without replacement from populations of ballot cards of which
some contain no valid vote, for risk limit $5\%$.
The populations contain 10,000, 100,000, or 500,000 ballot cards, of which 10\%, 25\%, 50\% or 75\% do not
contain a valid vote for either of two candidates under consideration.
The fraction of valid votes for the winner among valid votes is $\theta$. 
Votes are encoded using the SHANGRLA assorter for plurality contests:
A vote for the reported winner is ``1,'' a vote for the reported loser is ``0,'' and an invalid vote or vote for anyone else is ``1/2.''
The population mean is thus $\theta' := \theta(1-b) + b/2$, where $b$ is the fraction of ballot cards with no
vote for either of the two candidates.
For a given $\theta$, $\theta'$ shrinks as the percentage of ballot cards with no valid vote grows.
`apKelly' is \emph{a priori} Kelly.
The best result for each $\theta$, $N$, and percentage of non-votes is in bold font.
}
\end{table}

\begin{table}
\centering
\tiny
\begin{tabular}{lll|rrrr|rrrr|rrrr} 
& & & \multicolumn{4}{|c|}{$N=$10,000, \%blank} &  \multicolumn{4}{|c|}{$N=$100,000 \%blank} & \multicolumn{4}{|c}{$N=$500,000 \%blank} \\ 
$\theta$ & Method & params & 10 & 25 & 50 & 75  & 10 & 25 & 50 & 75  & 10 & 25 & 50 & 75  \\
\hline 0.6 & sqKelly & & 208  & 235  & 353  & 693  & 204  & 240  & 372  & 737  & 202  & 251  & 364  & 742  \\
\cline{2-15} & apKelly & $\eta=$0.51 & 841  & 995  & 1,456  & 2,685  & 888  & 1,047  & 1,577  & 3,105  & 874  & 1,063  & 1,565  & 3,159  \\
\cline{2-15}
& ALPHA & $\eta=$0.51 $d=$10 & 229  & 272  & 482  & 1,332  & 228  & 283  & 521  & 1,568  & 228  & 298  & 514  & 1,609  \\
&  & $\eta=$0.51 $d=$100 & 263  & 324  & 554  & 1,457  & 267  & 333  & 609  & 1,718  & 266  & 344  & 601  & 1,761  \\
&  & $\eta=$0.51 $d=$1000 & 417  & 516  & 854  & 1,947  & 446  & 554  & 959  & 2,426  & 439  & 565  & 954  & 2,508  \\
 &  & $\eta=$0.51 $d=\infty$ & 706  & 910  & 1,548  & 3,217  & 941  & 1,292  & 2,695  & 8,164  & 955  & 1,376  & 2,973  & 10,830  \\
\cline{2-15} & apKelly & $\eta=$0.52 & 458  & 538  & 795  & 1,510  & 471  & 553  & 835  & 1,643  & 464  & 561  & 833  & 1,666  \\
\cline{2-15}
& ALPHA & $\eta=$0.52 $d=$10 & 227  & 270  & 480  & 1,328  & 226  & 281  & 518  & 1,564  & 225  & 296  & 511  & 1,604  \\
&  & $\eta=$0.52 $d=$100 & 250  & 310  & 538  & 1,438  & 253  & 317  & 591  & 1,693  & 249  & 328  & 581  & 1,736  \\
&  & $\eta=$0.52 $d=$1000 & 348  & 439  & 764  & 1,848  & 362  & 463  & 846  & 2,273  & 361  & 472  & 845  & 2,358  \\
 &  & $\eta=$0.52 $d=\infty$ & 458  & 609  & 1,155  & 2,797  & 512  & 706  & 1,521  & 5,280  & 508  & 729  & 1,554  & 5,950  \\
\cline{2-15} & apKelly & $\eta=$0.55 & 227  & 261  & 385  & 771  & 225  & 265  & 409  & 804  & 222  & 274  & 402  & 817  \\
\cline{2-15}
& ALPHA & $\eta=$0.55 $d=$10 & 225  & 264  & 471  & 1,320  & 221  & 275  & 509  & 1,548  & 221  & 291  & 504  & 1,591  \\
&  & $\eta=$0.55 $d=$100 & 219  & 270  & 492  & 1,379  & 218  & 279  & 538  & 1,622  & 216  & 292  & 527  & 1,668  \\
&  & $\eta=$0.55 $d=$1000 & 232  & 299  & 563  & 1,583  & 234  & 306  & 618  & 1,880  & 229  & 317  & 610  & 1,934  \\
 &  & $\eta=$0.55 $d=\infty$ & 240  & 315  & 627  & 1,902  & 242  & 325  & 689  & 2,463  & 238  & 334  & 685  & 2,560  \\
\cline{2-15} & apKelly & $\eta=$0.6 & \bf{172}  & \bf{192}  & 296  & \bf{555}  & \bf{171}  & \bf{199}  & \bf{313}  & \bf{612}  & \bf{170}  & \bf{208}  & \bf{302}  & \bf{626}  \\
\cline{2-15}
& ALPHA & $\eta=$0.6 $d=$10 & 220  & 257  & 457  & 1,304  & 214  & 267  & 499  & 1,526  & 217  & 284  & 494  & 1,572  \\
&  & $\eta=$0.6 $d=$100 & 191  & 228  & 430  & 1,291  & 186  & 234  & 467  & 1,510  & 186  & 249  & 459  & 1,549  \\
&  & $\eta=$0.6 $d=$1000 & 176  & 209  & 393  & 1,240  & 173  & 217  & 422  & 1,423  & 173  & 225  & 413  & 1,453  \\
 &  & $\eta=$0.6 $d=\infty$ & 174  & 207  & 381  & 1,205  & 172  & 212  & 408  & 1,350  & \bf{170}  & 221  & 402  & 1,376  \\
\cline{2-15} & apKelly & $\eta=$0.7 & 598  & 586  & 793  & 1,093  & 1,681  & 1,695  & 2,361  & 3,234  & 3,653  & 5,025  & 5,917  & 7,400  \\
\cline{2-15}
& ALPHA & $\eta=$0.7 $d=$10 & 220  & 248  & 441  & 1,271  & 213  & 259  & 481  & 1,481  & 218  & 275  & 476  & 1,530  \\
&  & $\eta=$0.7 $d=$100 & 207  & 210  & 360  & 1,142  & 195  & 217  & 386  & 1,313  & 204  & 231  & 380  & 1,349  \\
&  & $\eta=$0.7 $d=$1000 & 270  & 219  & 301  & 871  & 263  & 230  & 318  & 937  & 268  & 246  & 306  & 966  \\
 &  & $\eta=$0.7 $d=\infty$ & 390  & 244  & \bf{295}  & 761  & 505  & 261  & \bf{313}  & 803  & 534  & 287  & \bf{302}  & 816  \\
\hline 0.7 & sqKelly & & 76  & 90  & 138  & 273  & 75  & 92  & 136  & 275  & 76  & 91  & 138  & 276  \\
\cline{2-15} & apKelly & $\eta=$0.51 & 418  & 500  & 742  & 1,421  & 427  & 512  & 765  & 1,535  & 426  & 512  & 770  & 1,550  \\
\cline{2-15}
& ALPHA & $\eta=$0.51 $d=$10 & 60  & 76  & 140  & 405  & 58  & 78  & 137  & 420  & 60  & 77  & 140  & 424  \\
&  & $\eta=$0.51 $d=$100 & 92  & 116  & 199  & 507  & 91  & 118  & 199  & 534  & 93  & 118  & 202  & 541  \\
&  & $\eta=$0.51 $d=$1000 & 184  & 231  & 381  & 862  & 192  & 245  & 400  & 962  & 193  & 245  & 406  & 981  \\
 &  & $\eta=$0.51 $d=\infty$ & 347  & 456  & 802  & 1,796  & 453  & 637  & 1,327  & 4,152  & 469  & 671  & 1,476  & 5,427  \\
\cline{2-15} & apKelly & $\eta=$0.52 & 217  & 260  & 391  & 763  & 219  & 265  & 393  & 789  & 219  & 264  & 396  & 796  \\
\cline{2-15}
& ALPHA & $\eta=$0.52 $d=$10 & 59  & 75  & 138  & 403  & 57  & 77  & 136  & 418  & 59  & 76  & 139  & 422  \\
&  & $\eta=$0.52 $d=$100 & 86  & 110  & 192  & 498  & 86  & 112  & 191  & 524  & 87  & 111  & 194  & 530  \\
&  & $\eta=$0.52 $d=$1000 & 152  & 196  & 340  & 814  & 155  & 205  & 354  & 902  & 157  & 206  & 358  & 917  \\
 &  & $\eta=$0.52 $d=\infty$ & 218  & 298  & 580  & 1,519  & 239  & 343  & 738  & 2,649  & 241  & 346  & 765  & 2,954  \\
\cline{2-15} & apKelly & $\eta=$0.55 & 95  & 114  & 174  & 340  & 94  & 116  & 172  & 348  & 96  & 115  & 175  & 350  \\
\cline{2-15}
& ALPHA & $\eta=$0.55 $d=$10 & 57  & 73  & 135  & 398  & 55  & 74  & 132  & 413  & 57  & 74  & 135  & 416  \\
&  & $\eta=$0.55 $d=$100 & 72  & 93  & 172  & 472  & 71  & 95  & 171  & 496  & 72  & 94  & 174  & 502  \\
&  & $\eta=$0.55 $d=$1000 & 94  & 127  & 246  & 683  & 94  & 131  & 253  & 741  & 95  & 131  & 256  & 753  \\
 &  & $\eta=$0.55 $d=\infty$ & 103  & 144  & 301  & 981  & 103  & 148  & 317  & 1,213  & 105  & 149  & 321  & 1,243  \\
\cline{2-15} & apKelly & $\eta=$0.6 & 56  & 67  & 102  & 201  & 55  & 68  & 101  & 201  & 56  & 67  & 103  & 202  \\
\cline{2-15}
& ALPHA & $\eta=$0.6 $d=$10 & 53  & 69  & 129  & 391  & 52  & 70  & 127  & 405  & 53  & 70  & 130  & 407  \\
&  & $\eta=$0.6 $d=$100 & 56  & 74  & 145  & 433  & 55  & 76  & 143  & 451  & 56  & 75  & 147  & 456  \\
&  & $\eta=$0.6 $d=$1000 & 59  & 80  & 165  & 523  & 58  & 82  & 164  & 553  & 59  & 81  & 167  & 559  \\
 &  & $\eta=$0.6 $d=\infty$ & 60  & 81  & 172  & 591  & 59  & 83  & 172  & 637  & 60  & 83  & 175  & 646  \\
\cline{2-15} & apKelly & $\eta=$0.7 & \bf{41}  & \bf{50}  & \bf{75}  & \bf{152}  & \bf{41}  & \bf{52}  & \bf{76}  & \bf{153}  & \bf{42}  & \bf{52}  & \bf{77}  & \bf{154}  \\
\cline{2-15}
& ALPHA & $\eta=$0.7 $d=$10 & 48  & 63  & 120  & 376  & 47  & 64  & 117  & 389  & 49  & 64  & 121  & 391  \\
&  & $\eta=$0.7 $d=$100 & 43  & 56  & 108  & 364  & 43  & 57  & 107  & 376  & 44  & 56  & 110  & 377  \\
&  & $\eta=$0.7 $d=$1000 & 42  & 53  & 103  & 342  & 42  & 55  & 101  & 353  & 43  & 54  & 103  & 356  \\
&  & $\eta=$0.7 $d=\infty$ & 42  & 53  & 102  & 336  & \bf{41}  & 55  & 100  & 347  & 43  & 54  & 103  & 349 
 \end{tabular}
\caption{\protect \label{tab:without-replacement-blanks-2}
Same as table \ref{tab:without-replacement-blanks-1} for other values of $\theta$.
}
\end{table}

Table~\ref{tab:summary} summarizes tables~\ref{tab:without-replacement-blanks-1} and \ref{tab:without-replacement-blanks-2},
using the geometric mean of the ratio of the average sample size to the 
average sample size of the best method for each combination of $\theta$, $N$, and percentage of blanks, a total
of 60 conditions.
ALPHA with $\eta=0.6$ and $d=100$ was best overall by that measure, with a geometric mean ratio of 1.54.
Several other parameter combinations in ALPHA performed similarly.

\begin{table}
\centering
\tiny
\begin{tabular}{llr}\\ 
Method & Parameters & Score \\
\hline SqKelly & & 1.89 \\ 
 \hline a priori Kelly 
 & $\eta=$0.51 & 2.90 \\
 & $\eta=$0.52 & 1.97 \\
 & $\eta=$0.55 & 2.14 \\
 & $\eta=$0.6 & 2.98 \\
 & $\eta=$0.7 & 7.49 \\
\hline ALPHA 
 & $\eta=$0.51 $d=$10 & 1.62 \\ 
 & $\eta=$0.51 $d=$100 & 1.77 \\ 
 & $\eta=$0.51 $d=$1000 & 2.29 \\ 
 & $\eta=$0.51 $d=\infty$ & 3.80 \\
\cline{2-3}
 & $\eta=$0.52 $d=$10 & 1.61 \\ 
 & $\eta=$0.52 $d=$100 & 1.73 \\ 
 & $\eta=$0.52 $d=$1000 & 2.08 \\ 
 & $\eta=$0.52 $d=\infty$ & 2.62 \\
\cline{2-3}
 & $\eta=$0.55 $d=$10 & 1.60 \\ 
 & $\eta=$0.55 $d=$100 & 1.63 \\ 
 & $\eta=$0.55 $d=$1000 & 1.71 \\ 
 & $\eta=$0.55 $d=\infty$ & 2.15 \\
\cline{2-3}
 & $\eta=$0.6 $d=$10 & 1.58 \\ 
 & $\eta=$0.6 $d=$100 & \bf{1.54} \\ 
 & $\eta=$0.6 $d=$1000 & 1.59 \\ 
 & $\eta=$0.6 $d=\infty$ & 2.40 \\
\cline{2-3}
 & $\eta=$0.7 $d=$10 & 1.57 \\ 
 & $\eta=$0.7 $d=$100 & 1.56 \\ 
 & $\eta=$0.7 $d=$1000 & 1.99 \\ 
 & $\eta=$0.7 $d=\infty$ & 3.90 \\
\end{tabular}
\caption{\protect \label{tab:summary}
Summary of tables~\ref{tab:without-replacement-blanks-1} and \ref{tab:without-replacement-blanks-2}:
the geometric mean of the ratio of the mean sample size for each method in each experimental condition to
that of the method with the smallest mean sample size for that condition.
The smallest is in bold font.
}
\end{table}

\subsection{Simulations for comparison audits} \label{sec:comparison-audit-sims}

For comparison audits, SHANGRLA assorters take nonnegative, bounded values, but generally have  several points of 
support, depending on the social choice function.
See section 3.2 of \citet{stark20}.
Again, the statistical null hypothesis is that the mean value of the assorter is not greater than 1/2, and rejecting that
hypothesis is evidence that the corresponding assertion is correct.

Several supermartingales can be used to test such a hypothesis from samples with or without replacement, including the ALPHA
supermartingales,
``Kaplan-Wald'' martingale \citep{stark09b}, the ``Kaplan-Kolmogorov'' martingale \citep{starkEvans19,stark20}, Kaplan's martingale 
\citep{starkEvans19,stark20},
and the martingales in \citet{waudby-smithRamdas21,waudby-smithEtal21}.
As mentioned above, the betting martingales in \citet{waudby-smithRamdas21,waudby-smithEtal21} are
identical to the ALPHA supermartingales but for how $\eta_i$ is chosen.
The Kaplan-Wald, Kaplan-Kolmogorov, and Kaplan martingales are also in this family of betting martingales.

As before, let $\theta_i$ denote the mean of the population just before the $j$th draw, if the null hypothesis is true.
For sampling with replacement, $\theta_i = \theta$, the hypothesized mean, and for sampling without replacement,
$\theta_i = (N\theta-S_{i-1})/(N-i+1)$.
The Kaplan-Wald martingale is $t_j := \prod_{k=1}^j \left (g(X_k/\theta_k - 1)+1 \right )$.
The tuning parameter $g \in [0, 1]$ does not affect the validity of the test, but hedges against the possibility that some $X_i = 0$.
Kaplan's martingale is the Kalpan-Wald martingale integrated with respect to $g$ over the interval $[0, 1]$.
The Kaplan-Kolmogorov martingale is $T_j := \prod_{i=1}^j (X_i+g)/(\theta_i+g)$, 
Again, the tuning parameter $g \ge 0$ does not affect the validity of the test, but hedges against the possibility that some $X_i = 0$.
It is straightforward to verify that under the null, these are nonnegative supermartingales with expected value 1.
Both of these can be written in the form \ref{eq:lambda-rilacs}.

For comparison audits, a reference alternative value $\eta$ for ALPHA and \emph{a priori} Kelly could be derived from assumptions about the 
frequency of errors of different types.
For instance, one might suppose that errors that turn votes from a reported loser 
into votes for a reported winner occur in about 1 in 1,000 ballot cards; errors that turn a valid vote for a loser into an 
undervote occur in about 1 in 100 ballot cards; etc.

To assess the relative performance of these supermartingales for comparison audits, they were applied to
pseudo-random samples from nonnegative populations that had mass $0.001$ at zero (corresponding to errors that overstate the
margin by the maximum possible, e.g., that erroneously interpreted a vote for the loser as a vote for the winner), 
mass $m \in \{0.01, 0.1, 0.25, 0.5, 0.75, 0.9, 0.99\}$
at $1$, and the remain mass uniformly distributed on $[0, 1]$.
The results are in table \ref{tab:comparison-1} for $m \in \{0.99, 0.9, 0.75\}$ and 
in table~\ref{tab:comparison-2} for $m \in \{0.25, 0.1, 0.01\}$,
for a variety of choices of $\eta$ for methods that use it, and a variety of choices of some of the other parameters
in the methods.
ALPHA is competitive.
Table~\ref{tab:comparison-summary} shows the geometric mean of the ratio of each method's average sample size
to the smallest average sample size for each combination of $m$ and $N$.
ALPHA with $\eta=0.9$ and $d=10$ had the lowest geometric mean ratio of all the methods tested,
and ALPHA with $\eta=0.99$ and $d=10$ was a close second.

\begin{table}
\centering
\tiny
\begin{tabular}{lll|rrr} 
 mass at 1 & method & params & $N=10{,}000$ &  $N=100{,}000$ & $N=500{,}000$  \\
\hline 0.99 & sqKelly & & 23  & 23  & 23  \\
\cline{2-6} & apKelly & $\eta=$0.99 & \bf{5}  & \bf{5}  & \bf{5}  \\
\cline{2-6}
& ALPHA & $\eta=$0.99 $d=$10 & \bf{5}  & \bf{5}  & \bf{5}  \\
& ALPHA & $\eta=$0.99 $d=$100 & \bf{5}  & \bf{5}  & \bf{5}  \\
\cline{2-6} & apKelly & $\eta=$0.9 & 6  & 6  & 6  \\
\cline{2-6}
& ALPHA & $\eta=$0.9 $d=$10 & \bf{5}  & \bf{5}  & \bf{5}  \\
& ALPHA & $\eta=$0.9 $d=$100 & 6  & 6  & 6  \\
\cline{2-6} & apKelly & $\eta=$0.75 & 8  & 8  & 8  \\
\cline{2-6}
& ALPHA & $\eta=$0.75 $d=$10 & 7  & 7  & 7  \\
& ALPHA & $\eta=$0.75 $d=$100 & 8  & 8  & 8  \\
\cline{2-6} & apKelly & $\eta=$0.55 & 32  & 32  & 32  \\
\cline{2-6}
& ALPHA & $\eta=$0.55 $d=$10 & 11  & 11  & 11  \\
& ALPHA & $\eta=$0.55 $d=$100 & 19  & 19  & 19  \\
\cline{2-6}
 & Kaplan-Kolmogorov & g=0.01 & \bf{5}  & \bf{5}  & \bf{5}  \\
 & Kaplan-Kolmogorov & g=0.1 & \bf{5}  & \bf{5}  & \bf{5}  \\
 & Kaplan-Kolmogorov & g=0.2 & 6  & 6  & 6  \\
\cline{2-6}
 & Kaplan-Wald & g=0.99 & \bf{5}  & \bf{5}  & \bf{5}  \\
 & Kaplan-Wald & g=0.9 & \bf{5}  & \bf{5}  & \bf{5}  \\
 & Kaplan-Wald & g=0.8 & 6  & 6  & 6  \\
\hline 0.90 & sqKelly & & 26  & 26  & 26  \\
\cline{2-6} & apKelly & $\eta=$0.99 & \bf{6}  & \bf{6}  & \bf{6}  \\
\cline{2-6}
& ALPHA & $\eta=$0.99 $d=$10 & \bf{6}  & \bf{6}  & \bf{6}  \\
& ALPHA & $\eta=$0.99 $d=$100 & \bf{6}  & \bf{6}  & \bf{6}  \\
\cline{2-6} & apKelly & $\eta=$0.9 & 7  & 7  & 7  \\
\cline{2-6}
& ALPHA & $\eta=$0.9 $d=$10 & \bf{6}  & \bf{6}  & \bf{6}  \\
& ALPHA & $\eta=$0.9 $d=$100 & 7  & 7  & 7  \\
\cline{2-6} & apKelly & $\eta=$0.75 & 9  & 9  & 9  \\
\cline{2-6}
& ALPHA & $\eta=$0.75 $d=$10 & 8  & 8  & 8  \\
& ALPHA & $\eta=$0.75 $d=$100 & 9  & 9  & 9  \\
\cline{2-6} & apKelly & $\eta=$0.55 & 35  & 36  & 36  \\
\cline{2-6}
& ALPHA & $\eta=$0.55 $d=$10 & 12  & 12  & 12  \\
& ALPHA & $\eta=$0.55 $d=$100 & 21  & 22  & 22  \\
\cline{2-6}
 & Kaplan-Kolmogorov & g=0.01 & \bf{6}  & \bf{6}  & \bf{6}  \\
 & Kaplan-Kolmogorov & g=0.1 & \bf{6}  & \bf{6}  & \bf{6}  \\
 & Kaplan-Kolmogorov & g=0.2 & 7  & 7  & 7  \\
\cline{2-6}
 & Kaplan-Wald & g=0.99 & \bf{6}  & \bf{6}  & \bf{6}  \\
 & Kaplan-Wald & g=0.9 & \bf{6}  & \bf{6}  & \bf{6}  \\
 & Kaplan-Wald & g=0.8 & 7  & 7  & 7  \\
\hline 0.75 & sqKelly & & 32  & 31  & 31  \\
\cline{2-6} & apKelly & $\eta=$0.99 & \bf{8}  & \bf{8}  & \bf{8}  \\
\cline{2-6}
& ALPHA & $\eta=$0.99 $d=$10 & \bf{8}  & \bf{8}  & \bf{8}  \\
& ALPHA & $\eta=$0.99 $d=$100 & \bf{8}  & \bf{8}  & \bf{8}  \\
\cline{2-6} & apKelly & $\eta=$0.9 & \bf{8}  & \bf{8}  & \bf{8}  \\
\cline{2-6}
& ALPHA & $\eta=$0.9 $d=$10 & \bf{8}  & \bf{8}  & \bf{8}  \\
& ALPHA & $\eta=$0.9 $d=$100 & \bf{8}  & \bf{8}  & \bf{8}  \\
\cline{2-6} & apKelly & $\eta=$0.75 & 11  & 11  & 11  \\
\cline{2-6}
& ALPHA & $\eta=$0.75 $d=$10 & 10  & 10  & 10  \\
& ALPHA & $\eta=$0.75 $d=$100 & 11  & 11  & 11  \\
\cline{2-6} & apKelly & $\eta=$0.55 & 43  & 43  & 43  \\
\cline{2-6}
& ALPHA & $\eta=$0.55 $d=$10 & 16  & 16  & 16  \\
& ALPHA & $\eta=$0.55 $d=$100 & 27  & 27  & 27  \\
\cline{2-6}
 & Kaplan-Kolmogorov & g=0.01 & \bf{8}  & \bf{8}  & \bf{8}  \\
 & Kaplan-Kolmogorov & g=0.1 & \bf{8}  & \bf{8}  & \bf{8}  \\
 & Kaplan-Kolmogorov & g=0.2 & 9  & 9  & 9  \\
\cline{2-6}
 & Kaplan-Wald & g=0.99 & \bf{8}  & \bf{8}  & \bf{8}  \\
 & Kaplan-Wald & g=0.9 & \bf{8}  & \bf{8}  & \bf{8}  \\
 & Kaplan-Wald & g=0.8 & \bf{8}  & \bf{8}  & \bf{8}  \\
\hline 0.50 & sqKelly & & 48  & 48  & 48  \\
\cline{2-6} & apKelly & $\eta=$0.99 & 16  & 16  & 16  \\
\cline{2-6}
& ALPHA & $\eta=$0.99 $d=$10 & 16  & 16  & 16  \\
& ALPHA & $\eta=$0.99 $d=$100 & 15  & 15  & \bf{15}  \\
\cline{2-6} & apKelly & $\eta=$0.9 & \bf{14}  & 15  & \bf{15}  \\
\cline{2-6}
& ALPHA & $\eta=$0.9 $d=$10 & 16  & 16  & 16  \\
& ALPHA & $\eta=$0.9 $d=$100 & 15  & 15  & \bf{15}  \\
\cline{2-6} & apKelly & $\eta=$0.75 & 18  & 18  & 18  \\
\cline{2-6}
& ALPHA & $\eta=$0.75 $d=$10 & 19  & 19  & 19  \\
& ALPHA & $\eta=$0.75 $d=$100 & 18  & 18  & 18  \\
\cline{2-6} & apKelly & $\eta=$0.55 & 65  & 65  & 65  \\
\cline{2-6}
& ALPHA & $\eta=$0.55 $d=$10 & 28  & 28  & 28  \\
& ALPHA & $\eta=$0.55 $d=$100 & 42  & 43  & 43  \\
\cline{2-6}
 & Kaplan-Kolmogorov & g=0.01 & 16  & 16  & 16  \\
 & Kaplan-Kolmogorov & g=0.1 & \bf{14}  & \bf{14}  & \bf{15}  \\
 & Kaplan-Kolmogorov & g=0.2 & 15  & 15  & \bf{15}  \\
\cline{2-6}
 & Kaplan-Wald & g=0.99 & 16  & 17  & 17  \\
 & Kaplan-Wald & g=0.9 & 15  & 15  & \bf{15}  \\
 & Kaplan-Wald & g=0.8 & \bf{14}  & 15  & \bf{15}  
\end{tabular} 

\caption{\protect \label{tab:comparison-1} Mean sample sizes to reject the hypothesis that the mean is less than
or equal to $1/2$ at significance level $0.05$ for various methods, in 10,000  simulations with mass 0.001 zero, mass $m$ at 1, and mass $1-m-0.001$ uniformly  distributed on $[0, 1]$, for values of $m$ between 0.99 and 0.5. 
The smallest mean sample size for each combination of $m$ and $N$ is in bold font.
}
\end{table}
 
\begin{table}
\centering
\tiny
\begin{tabular}{lll|rrr} 
 mass at 1 & method & params & $N=10{,}000$ &  $N=100{,}000$ & $N=500{,}000$  \\
\hline 0.25 & sqKelly & & 107  & 104  & 104  \\
\cline{2-6} & apKelly & $\eta=$0.99 & 645  & 3,970  & 71,802  \\
\cline{2-6}
& ALPHA & $\eta=$0.99 $d=$10 & 65  & 61  & 61  \\
& ALPHA & $\eta=$0.99 $d=$100 & 77  & 72  & 72  \\
\cline{2-6} & apKelly & $\eta=$0.9 & 81  & 75  & 74  \\
\cline{2-6}
& ALPHA & $\eta=$0.9 $d=$10 & 62  & 58  & 59  \\
& ALPHA & $\eta=$0.9 $d=$100 & 57  & 53  & 53  \\
\cline{2-6} & apKelly & $\eta=$0.75 & \bf{50}  & \bf{47}  & \bf{47}  \\
\cline{2-6}
& ALPHA & $\eta=$0.75 $d=$10 & 67  & 63  & 63  \\
& ALPHA & $\eta=$0.75 $d=$100 & 52  & 49  & 49  \\
\cline{2-6} & apKelly & $\eta=$0.55 & 138  & 134  & 135  \\
\cline{2-6}
& ALPHA & $\eta=$0.55 $d=$10 & 89  & 84  & 85  \\
& ALPHA & $\eta=$0.55 $d=$100 & 107  & 103  & 104  \\
\cline{2-6}
 & Kaplan-Kolmogorov & g=0.01 & 900  & 6,388  & 109,384  \\
 & Kaplan-Kolmogorov & g=0.1 & 99  & 92  & 92  \\
 & Kaplan-Kolmogorov & g=0.2 & 61  & 56  & 56  \\
\cline{2-6}
 & Kaplan-Wald & g=0.99 & 1,163  & 9,432  & 127,920  \\
 & Kaplan-Wald & g=0.9 & 189  & 223  & 240  \\
 & Kaplan-Wald & g=0.8 & 82  & 76  & 74  \\
\hline 0.10 & sqKelly & & 366  & 320  & 333  \\
\cline{2-6} & apKelly & $\eta=$0.99 & 7,713  & 75,917  & 383,563  \\
\cline{2-6}
& ALPHA & $\eta=$0.99 $d=$10 & 507  & 426  & 453  \\
& ALPHA & $\eta=$0.99 $d=$100 & 886  & 756  & 821  \\
\cline{2-6} & apKelly & $\eta=$0.9 & 5,807  & 66,399  & 339,357  \\
\cline{2-6}
& ALPHA & $\eta=$0.9 $d=$10 & 463  & 387  & 413  \\
& ALPHA & $\eta=$0.9 $d=$100 & 575  & 472  & 511  \\
\cline{2-6} & apKelly & $\eta=$0.75 & 1,007  & 1,370  & 4,471  \\
\cline{2-6}
& ALPHA & $\eta=$0.75 $d=$10 & 449  & 377  & 400  \\
& ALPHA & $\eta=$0.75 $d=$100 & \bf{354}  & \bf{286}  & \bf{308}  \\
\cline{2-6} & apKelly & $\eta=$0.55 & 418  & 371  & 384  \\
\cline{2-6}
& ALPHA & $\eta=$0.55 $d=$10 & 513  & 437  & 462  \\
& ALPHA & $\eta=$0.55 $d=$100 & 480  & 413  & 434  \\
\cline{2-6}
 & Kaplan-Kolmogorov & g=0.01 & 7,849  & 75,963  & 383,609  \\
 & Kaplan-Kolmogorov & g=0.1 & 6,429  & 68,694  & 350,602  \\
 & Kaplan-Kolmogorov & g=0.2 & 4,813  & 51,244  & 304,399  \\
\cline{2-6}
 & Kaplan-Wald & g=0.99 & 7,877  & 76,362  & 385,749  \\
 & Kaplan-Wald & g=0.9 & 7,134  & 72,541  & 367,628  \\
 & Kaplan-Wald & g=0.8 & 6,105  & 66,399  & 339,357  \\
\hline 0.01 & sqKelly & & \bf{7,554}  & 51,287  & 233,857  \\
\cline{2-6} & apKelly & $\eta=$0.99 & 9,473  & 94,357  & 472,671  \\
\cline{2-6}
& ALPHA & $\eta=$0.99 $d=$10 & 8,490  & 38,766  & 57,971  \\
& ALPHA & $\eta=$0.99 $d=$100 & 9,263  & 66,629  & 149,404  \\
\cline{2-6} & apKelly & $\eta=$0.9 & 9,454  & 94,060  & 470,344  \\
\cline{2-6}
& ALPHA & $\eta=$0.9 $d=$10 & 8,365  & 35,905  & 51,521  \\
& ALPHA & $\eta=$0.9 $d=$100 & 9,061  & 57,401  & 111,034  \\
\cline{2-6} & apKelly & $\eta=$0.75 & 9,342  & 93,102  & 447,670  \\
\cline{2-6}
& ALPHA & $\eta=$0.75 $d=$10 & 8,280  & 32,919  & 45,307  \\
& ALPHA & $\eta=$0.75 $d=$100 & 8,457  & 41,181  & 64,860  \\
\cline{2-6} & apKelly & $\eta=$0.55 & 7,627  & 56,930  & 270,663  \\
\cline{2-6}
& ALPHA & $\eta=$0.55 $d=$10 & 8,342  & 31,804  & 42,753  \\
& ALPHA & $\eta=$0.55 $d=$100 & 8,201  & \bf{30,688}  & \bf{41,143}  \\
\cline{2-6}
 & Kaplan-Kolmogorov & g=0.01 & 9,474  & 94,377  & 472,671  \\
 & Kaplan-Kolmogorov & g=0.1 & 9,453  & 94,040  & 470,641  \\
 & Kaplan-Kolmogorov & g=0.2 & 9,444  & 94,031  & 467,868  \\
\cline{2-6}
 & Kaplan-Wald & g=0.99 & 9,478  & 94,357  & 472,770  \\
 & Kaplan-Wald & g=0.9 & 9,460  & 94,050  & 471,581  \\
 & Kaplan-Wald & g=0.8 & 9,454  & 94,060  & 470,344  \\
 \end{tabular}
 \caption{\protect \label{tab:comparison-2} Same as table \ref{tab:comparison-1} for values of
$m$ between 0.25 and 0.01.}
\end{table}

\begin{table}
\centering
\tiny
\begin{tabular}{llr}\\ 
Method & Parameters & Score \\
\hline SqKelly & & 2.82 \\ 
 \hline a priori Kelly 
 & $\eta=$0.99 & 5.11 \\
 & $\eta=$0.9 & 2.81 \\
 & $\eta=$0.75 & 1.89 \\
 & $\eta=$0.55 & 3.62 \\
\hline ALPHA 
 & $\eta=$0.99 $d=$10 & 1.16 \\ 
 & $\eta=$0.99 $d=$100 & 1.37 \\ 
\cline{2-3}
 & $\eta=$0.9 $d=$10 & \bf{1.14} \\ 
 & $\eta=$0.9 $d=$100 & 1.27 \\ 
\cline{2-3}
 & $\eta=$0.75 $d=$10 & 1.31 \\ 
 & $\eta=$0.75 $d=$100 & 1.29 \\ 
\cline{2-3}
 & $\eta=$0.55 $d=$10 & 1.76 \\ 
 & $\eta=$0.55 $d=$100 & 2.41 \\ 
\cline{2-3}
\hline Kaplan-Kolmogorov
 & $g=$0.01 & 5.42\\ 
 & $g=$0.1 & 2.77\\ 
 & $g=$0.2 & 2.66\\ 
\hline Kaplan-Wald
 & $g=$0.99 & 5.67\\ 
 & $g=$0.9 & 3.14\\ 
 & $g=$0.8 & 2.81\\ 
\end{tabular}
\caption{\protect \label{tab:comparison-summary}
Summary of tables \ref{tab:comparison-1} and \ref{tab:comparison-2}: geometric mean of
the ratio of the average sample size to the smallest average sample size across values of $m$ and $N$.
Overall, the most efficient method (by this measure) is ALPHA with $\eta=0.9$ and $d=10$ (displayed in bold font).}
\end{table}

\section{Discussion}

\subsection{Non-adaptive ALPHA versus BRAVO}
BRAVO works with the conditional probability that a vote is for $w$, given that it is for $w$ or $\ell$,
using sampling with replacement.
That amounts to ignoring ballot cards that have valid votes for other candidates or that do not have a valid 
vote in the contest.
 ALPHA reduces to BRAVO in that situation, but because ALPHA can handle non-binary
values, it can also work with the unconditional population mean instead of ignoring those ballot cards.
In particular, if such ballot cards are assigned the value $1/2$ as in SHANGRLA, we can still audit by testing the
null hypothesis $\theta \le 1/2$.
Suppose we mimic BRAVO in every other respect: the sample is drawn with replacement, $u=1$, 
$\mu = 1/2 = \mu_i$ for all $i$, and $\eta_0 = \eta_i$ for all $i$.
What happens when we draw a ballot that does not contain a vote for $w$ or $\ell$, i.e., if $X_i = 1/2$?
The value of $T_i$ is the value of $T_{i-1}$ multiplied by
\begin{equation}
   \frac{1}{2} \cdot \frac{\eta_i}{1/2} + \frac{1}{2} \cdot \frac{1-\eta_i}{1/2} =  1.
\end{equation}
It follows that if, instead of ignoring ballot cards that do not have a valid vote for $w$ or for $\ell$, we treat such ballot cards as
$1/2$ in equation~\ref{eq:alphaBRAVOdef}, the resulting test is identical to BRAVO, with one difference: the value of $\eta$
corresponding to the reported results.
For BRAVO, $\eta = N_w/(N_w+N_\ell)$, while for the SHANGRLA assorter,  $\eta = (N_w + (N-N_w-N_\ell)/2)/N \le N_w/(N_w+N_\ell)$.

Non-adaptive ALPHA for sampling without replacement in a two-candidate contest with no invalid votes is equivalent
to Wald's SPRT for the population mean using sampling without replacement from a binary population.

\subsection{Other studies of ballot-polling RLA sample sizes}
There have been comparisons of ballot-polling sample sizes in the simplest case:
two-candidate plurality contests with no invalid votes.
For instance,
\citet{huangEtal20} compare previous methods for ballot-polling audits, including BRAVO, ClipAudit \citep{rivest17}, the Kaplan martingale \citep{stark20}, 
Kaplan-Wald \citep{stark09b,stark20}, Kaplan-Markov \citep{stark09b,stark20}, and Bayesian audits \citep{rivestShen12,rivest18} 
(calibrated to be risk limiting).
Similarly, \citet{waudby-smithEtal21} compare several martingale-based methods (including BRAVO), some of which
rely on the reported results, and some of which do not.

\subsection{Round-by-round ballot-polling RLAs}
In practice, ballot cards are not selected and inspected one at a time in RLAs.
(That strategy might require retrieving and opening the same storage container of ballot cards repeatedly, for instance, and it does not allow
multiple teams to work in parallel.)
Instead, for logistical efficiency, an initial sample is drawn that is expected to be large enough to confirm the outcome if the 
reported results are approximately correct.
Those ballot cards are retrieved and examined.
If they do not suffice to confirm the results (if the measured risk is larger than the risk limit), 
the sample is expanded by an amount that is expected to be large enough
to confirm the outcome, and so on.
(The auditors can decide to conduct a full hand count at any point in the process, rather than continuing to sample.)
Thus, individual ballot-by-ballot sequential validity may not be required.
Indeed, the first RLA methods did not use sequentially valid tests \citep{stark08a,stark09a},
instead prescribing a schedule of ``round sizes,'' and spending the total Type~I error budget across rounds.

\citet{zagorskiEtal21} note that this ``round-by-round'' sampling structure might make it possible to have tests that 
use smaller samples than tests that ensure ballot-by-ballot sequentially validity, and in particular smaller sample sizes than BRAVO
requires.
They show that in a two-candidate plurality contest with no invalid votes or non-votes, that is indeed
possible.
However, their method, Minerva, requires the sample to be IID Bernoulli:
it only applies to  two-candidate plurality contests with no invalid votes or non-votes.
It is not clear that it can implemented efficiently in real elections, because the number of ballot cards that contain a vote for the winner or the
loser in a random sample of a given size cannot be predicted in advance: some ballot cards have votes for other candidates or no valid vote in the contest or do not contain the contest.
Nor is the method conducive to auditing contests with more than two candidates or more than one contest at a time.

To see why, suppose that the first round is intended to contain $n_1$ ballot cards that have a valid vote either
for candidate $w$ or candidate $\ell$.  
How can auditors draw a random sample that guarantees that will happen, when some ballot cards do not contain
the contest, when there are invalid votes, and when there are other candidates in the contest?
If they draw a sample of size $n_1$, they will get a random number $N_1 \le n_1$ of cards that contain a valid vote either
for $w$ or for $\ell$, depending on the luck of the draw.
If they draw a sample large enough to have a large chance that $N_1 \ge n_1$, examining that larger sample could easily 
offset any savings from forfeiting ballot-by-ballot validity, if the proportion of ballot cards with a valid vote for $w$ or $\ell$ is
small.
Moreover, there is still some chance that $N_1 < n_1$, and another round of sampling will need to happen before the attained risk can be calculated.
And if $N_1 > n_1$, the $N_1-n_1$ ``extras'' cannot be used in the risk calculation in that round, because the round size is pre-specified.
If many (winner, loser) pairs are to be audited using the same sample, or if more than one contest is to be audited
using the same sample, the problem is exacerbated.

It is an open question whether there is a round-by-round method that can accommodate non-votes and votes
for other candidates and is more efficient than the methods in \citet{waudby-smithEtal21,stark20} and here.
It might be possible to maximize the $P$-value over a nuisance parameter (the number of non-votes and votes
for other candidates in the population), as in SUITE \citep{ottoboniEtal18}, or to use the SHANGRLA assorter
for plurality contests, which takes into account ballot cards with no valid vote in the contest and ballot cards with
a vote for other candidates in the contest, but Minerva does not have an obvious extension in either direction
because it assumes that the population mean by itself determines the probability distribution of
the sample.
While that is true for binary populations, it is not true when the population contains more than two values, e.g., the value $1/2$ that
SHANGRLA assorters assign to ballot cards with no valid vote in the contest, in addition to the values $0$ and $1$.

\subsection{Stratification}
As discussed in section~\ref{sec:stratified}, ALPHA and other test supermartingales offer a great deal of flexibility to
choose stratum selectors that
adaptively optimize union-intersection tests to increase their power.
Preliminary results in \cite{spertusStark22} suggest that this can reduce $P$-values by an order of magnitude
compared to previous methods, for the same sample size.

\subsection{Future work}
In many tests herein---two-candidate plurality contests with some invalid ballot cards or votes for other candidates, using
sampling with or without replacement, and ballot-level comparison audits---ALPHA with a shrinkage and truncation 
estimator is competitive with other methods, on average having the smallest sample size across a range of parameters.
It would be interesting
to explore a broader variety of estimates of $\theta_i$ based on $\eta$ and $X^{i-1}$ and their operating characteristics.
\citet{spertusStark22} studies the efficiency of some simple adaptive stratum selectors for stratified sampling,
as sketched in section~\ref{sec:stratified}.
It would be interesting to study the efficiency of ALPHA for batch-level comparison audits.
There are few competing methods that work so generally and guarantee sequential validity: 
Kaplan-Wald, Kaplan-Markov, and Kaplan's martingale \citep{stark09a,stark20},
and the betting martingales in \citet{waudby-smithEtal21}, as examined in sections~\ref{sec:blanks} and \ref{sec:comparison-audit-sims}.
It would also be interesting to explore the relative efficiency of batch-level comparison audits and ballot-polling audits for
a range of margins, batch-level vote distributions, and reporting errors.

\section{Conclusions}
BRAVO is based on Wald's sequential probability ratio test for $p$ from IID $\Bern(p)$ observations, for a simple (i.e., ``point'') null hypothesis against a simple alternative.
The SPRT for the Bernoulli distribution can easily be generalized in a way that has a number of advantages:
\begin{itemize}
   \item in situations where BRAVO can be applied, it can be tuned to perform comparably to BRAVO when the reported vote shares are correct, 
             and to perform far better than BRAVO when the reported vote shares are incorrect but the reported winner(s) really won
   \item it works for sampling with and without replacement and for Bernoulli sampling
   \item it can be used with stratified sampling, and has more power than SUITE \citep{ottoboniEtal18} in numerical experiments \citep{spertusStark22}
   \item it works for populations that are not binary, but merely bounded, allowing it to test any SHANGRLA assertion,
            including assertions for ballot-polling and ballot-level comparison audits
   \item it can be applied to batch-polling and batch-level comparison audits, sampling with and without replacement
   \item it works for batch-polling and batch-level comparison audits using sampling weights
   \item in simulations, its expected sample sizes are competitive with those of all known methods, for ballot polling with and without replacement and 
            for ballot-level comparison audits
   \end{itemize}
This generalization, ALPHA, tests the hypothesis that the mean of a finite, bounded population does not exceed a threshold.
It has a great deal of freedom to be optimized for different situations, parametrized by
estimators of the population mean after the $j$th sample has been drawn.
It can also accommodate sampling units that are batches rather than individuals, and sampling such batches
with or without replacement, with or without weights.
ALPHA is computationally efficient, far faster than some competing methods, such as the Kaplan martingale \citep{starkEvans19,stark20}.
Its statistical performance is competitive with that of the betting martingales introduced for RLAs in \citet{waudby-smithEtal21},
better against some alternatives and worse against others.
For comparison audits, it improves substantially on the Kaplan-Wald and Kaplan-Kolmogorov methods.
Like the Kaplan-Wald \citep{stark09b,stark20}, Kaplan-Kolmogorov \citep{stark20}, Kaplan martingale \citep{starkEvans19,stark20}, 
RiLACS \citep{waudby-smithEtal21}, and BRAVO \citep{lindemanEtal12}, it is based on Ville's inequality for nonnegative supermartingales \citep{ville39}.
Unlike all of those except some flavors of RiLACS, it adapts to the audit data, leading to increased power when the reported vote shares
are wrong but the reported outcomes are correct.
Overall, in the simulations involving sampling without replacement when some ballot cards do not contain a valid
vote, ALPHA with the truncated shrinkage estimator using $\eta=0.6$ and $d=100$ perfomed best, as measured by the
geometric mean of the ratios between the mean sample sizes and the best mean sample size, across conditions.
In simulations involving sampling without replacement from populations that correspond to ballot-level comparison audits,
ALPHA with the truncated shrinkage estimator using $\eta=0.9$ and $d=10$ performed best by the same measure.
A reference Python implementation is available at \url{https://github.com/pbstark/alpha}.

\begin{acks}[Acknowledgments] 
I am grateful to Andrew Appel, Amanda Glazer, Aaditya Ramdas, Jacob Spertus, Damjan Vukcevic, and Ian Waudby-Smith for comments on earlier drafts.
\end{acks}

\bibliographystyle{imsart-nameyear}
\bibliography{bib}

\end{document}